\documentclass[aps,prd,twocolumn,floatfix,nofootinbib,showpacs,superscriptaddress]{revtex4-1}

\usepackage[mathscr,scaled=1.15]{urwchancal}
\DeclareFontFamily{OT1}{pzc}{}
\DeclareFontShape{OT1}{pzc}{m}{it}%
{<-> s * [1.15] pzcmi7t}{}
\DeclareMathAlphabet{\mathpzc}{OT1}{pzc}{m}{it}

\usepackage{amsmath,amsfonts,amssymb,bm}
\usepackage{graphicx}
\usepackage{color}

\definecolor{purple}{rgb}{0.5,0,0.5}
\definecolor{blue}{rgb}{0.0,0,0.9}

\begin{document}

\title{Parton distribution amplitudes of light vector mesons}
\email[Communicating authors: ]{yxliu@pku.edu.cn, cdroberts@anl.gov}

\author{Fei Gao}
\affiliation{Department of Physics and State
Key Laboratory of Nuclear Physics and Technology, Peking
University, Beijing 100871, China}
\affiliation{Collaborative Innovation Center of Quantum Matter, Beijing 100871, China}

\author{Lei Chang}
\affiliation{CSSM, School of Chemistry and Physics
University of Adelaide, Adelaide SA 5005, Australia}

\author{Yu-Xin Liu }
\affiliation{Department of Physics and State Key Laboratory of
Nuclear Physics and Technology, Peking University, Beijing 100871,
China}
\affiliation{Collaborative Innovation Center of Quantum Matter, Beijing 100871, China}
\affiliation{Center for High Energy Physics, Peking University, Beijing 100871, China}

\author{Craig D.~Roberts}
\affiliation{Physics Division, Argonne National Laboratory, Argonne, IL 60439, USA}

\author{Sebastian M.~Schmidt}
\affiliation{Institute for Advanced Simulation, Forschungszentrum J\"ulich and JARA, D-52425 J\"ulich, Germany}

\date{30 April 2014}

\begin{abstract}
A rainbow-ladder truncation of QCD's Dyson-Schwinger equations is used to calculate $\rho$- and $\phi$-meson valence-quark (twist-two parton) distribution amplitudes (PDAs) via a light-front projection of their Bethe-Salpeter wave functions, which possess $S$- and $D$-wave components of comparable size in the meson rest frame.  All computed PDAs are broad concave functions, whose dilation with respect to the asymptotic distribution is an expression of dynamical chiral symmetry breaking.  The PDAs can be used to define an ordering of valence-quark light-front spatial-extent within mesons: this size is smallest within the pion and increases through the $\perp$-polarisation to the $\|$-polarisation of the vector mesons; effects associated with the breaking of SU$(3)$-flavour symmetry are significantly smaller than those associated with altering the polarisation of vector mesons.  Notably, the predicted pointwise behaviour of the $\rho$-meson PDAs is in quantitative agreement with that inferred recently via an analysis of diffractive vector-meson photoproduction experiments.
\end{abstract}

\pacs{
14.40.Be,    
11.10.St,    
12.38.Lg,	
13.60.Le	
\hspace*{\fill}\emph{Preprint no}. ADP-14-12/T870
}

\maketitle

\section{Introduction}
The properties of pseudoscalar mesons constituted from light quarks are strongly influenced by dynamical chiral symmetry breaking (DCSB), an emergent phenomenon in the Standard Model which plays a major role in the origin of more than 98\% of the visible mass in the Universe \cite{national2012Nuclear}.  For example, it is DCSB that explains both the momentum evolution of QCD's Lagrangian current-quark masses  \cite{Bhagwat:2003vw,Bhagwat:2006tu,Bowman:2005vx}, so that they acquire values at infrared momenta which are enhanced by two orders-of-magnitude, to values commensurate with those used in describing the baryon spectrum with constituent-quark models: $m \sim 4\,$MeV\,$\to\,M\sim 400\,$MeV; and also the quadratic increase of pseudoscalar meson masses with increasing current-quark mass \cite{GellMann:1968rz,Maris:1997hd,Chang:2011mu}: $m_{0^-}^2\propto m$.  Given that quantum mechanical models describe vector mesons as merely spin-flip excitations of pseudoscalar mesons, in analogy with para- ($\,^1\!S_0$) and ortho- ($\,^3\!S_1$) positronium, it is natural to enquire into the impact of DCSB within these companion vector-meson bound states.

A clean way to expose differences between the impact of DCSB in pseudoscalar and vector mesons is to compare their wave functions.  However, since DCSB is an emergent feature of quantum field theory and is strictly impossible in quantum mechanics with a finite number of degrees-of-freedom, such a comparison is not generally possible.  The closest thing one has to a wave function in quantum field theory is a meson's Poincar\'e-covariant Bethe-Salpeter wave function, $\psi_{\rm BS}$, which reduces to a Schr\"odinger wave function whenever a nonrelativistic limit is sensible \cite{Salpeter:1951sz}; but that is never the case for the valence-quark constituents of light mesons.

An answer to the problem of defining an appropriate quantum field theory wave function lies in projecting $\psi_{\rm BS}$ onto the light front.  The light-front wave function of an interacting quantum system provides a connection between dynamical properties of the underlying relativistic quantum field theory and notions familiar from nonrelativistic quantum mechanics \cite{Keister:1991sb,Coester:1992cg,Brodsky:1997de}.  It can translate features that arise purely through the infinitely-many-body nature of relativistic quantum field theory into images whose interpretation is seemingly more straightforward.  Herein, therefore, we provide a comparison and interpretation of light-front projections of pseudoscalar and vector meson Bethe-Salpeter wave functions; namely, their respective valence-quark (twist-two parton) distribution amplitudes (PDAs).

As wave functions, the PDAs of pseudoscalar- and vector-mesons are not directly measurable.  Nevertheless, many of their features can be constrained by experiment.  For example, the pion's valence-quark PDA modulates the momentum-dependence of its elastic electromagnetic form factor at large momentum transfers \cite{Efremov:1979qk,Lepage:1979zb,Lepage:1980fj,Farrar:1979aw}, a feature which is evident for $Q^2\gtrsim 8\,$GeV$^2$ \cite{Maris:1998hc,Chang:2013nia}; and vector meson PDAs can be inferred \cite{Forshaw:2003ki,Forshaw:2010py} from diffractive vector-meson production experiments, such as those performed at the Hadron Electron Ring Accelerator (HERA) \cite{Adloff:1997sc,Breitweg:1998gc,Chekanov:2007zr,Aaron:2009xp}.  More generally, in fact, meson PDAs can be used to express the cross-sections for numerous hard exclusive processes
\cite{Lepage:1979zb,Farrar:1979aw,Lepage:1980fj,Efremov:1979qk,Brodsky:1999ay,
Beneke:1999br,Beneke:2000ry,Beneke:2001ev}; and therefore the estimation of meson PDAs has long been topical.  In this context, complementing work by other authors using different tools (e.g., Refs.\,\cite{Ball:1996tb,Ball:1998sk,Bakulev:1998pf,Pimikov:2013usa,Dorokhov:2006xw,Choi:2007yu,%
Forshaw:2012im,Ahmady:2012dy,Braun:2003jg,Jansen:2009hr,Braun:2007zr,Arthur:2010xf}), herein we present both predictions for the $\rho$- and $\phi$-meson valence-quark PDAs, and their comparison with the pion's PDA, obtained using the same Dyson-Schwinger equation (DSE) framework employed recently to explain the electromagnetic pion form factor \cite{Chang:2013nia}.

This manuscript is arranged as follows.  In Sec.\,\ref{secVPDA} we describe formulae necessary to compute and understand our results for meson Bethe-Salpeter amplitudes and PDAs.  Those results are detailed in Sec.\,\ref{secResults}; and Sec.\,\ref{secEpilogue} provides a summary and perspective.

\section{Parton Distribution Amplitudes and Bethe-Salpeter wave functions}
\label{secVPDA}
\subsection{Twist-two PDAs}
In this section we explicitly develop formulae for the case of the $\rho$-meson.  Since the $\phi$ is also a $J^{PC}=1^{--}$ state, the only changes for that case are straightforward, being associated with replacing the $u/d$-quarks (we assume isospin symmetry) by the $s$-quarks and adjusting the meson mass: $m_\rho^2 \to m_\phi^2$.

The light-cone valence-quark (twist-two parton) distribution amplitudes for a $\rho$-meson with total momentum $P$ and helicity $\lambda$ may be defined as follows \cite{Ball:1996tb}:
\begin{subequations}
\label{rhoPDAs}
\begin{eqnarray}
\nonumber
\lefteqn{\langle 0|\bar{u}(-z/2)i \gamma_\mu d(z/2)|\rho^+(P,\lambda)\rangle
}\\
\nonumber
&=& P_\mu \frac{n\cdot \varepsilon^\lambda}{n\cdot P}
f_\rho m_\rho \int^1_0dx \, {\rm e}^{-ixn\cdot P}\,
\left[\varphi_\|(x,\zeta) - g^{(v)}_\bot(x,\zeta)\right]\\
&& \quad + \,\varepsilon^\lambda_\mu f_\rho m_\rho\int^1_0dx \, {\rm e}^{-ix n\cdot P}\, g^{(v)}_\bot(x,\zeta)\,,\\
\nonumber
\lefteqn{\langle 0|\bar{u}(-z/2)i\sigma_{\mu\nu}d(z/2)|\rho^+(P,\lambda)\rangle
}\\
&=& (\varepsilon^\lambda_\mu P_\nu-\varepsilon^\lambda_\nu P_\mu)f^\bot_\rho \int^1_0dx\,
{\rm e}^{-ix n\cdot P}\,\varphi_\bot(x,\zeta)\,, \\
\nonumber
\lefteqn{\langle 0|\bar{u}(-z/2)\gamma_\mu\gamma_5 d(z/2)|\rho^+(P,\lambda)\rangle }\\
 &=& \frac{i}{4} \epsilon _{\mu\nu\tau\sigma} \varepsilon^{\lambda}_\nu P_\tau n_\sigma f_\rho m_\rho \int^1_0 dx \, {\rm e}^{-ix n\cdot P}g^{(a)}_\bot(x,\zeta)   \,,
\end{eqnarray}
\end{subequations}
where:\footnote{We use a Euclidean metric:  $\{\gamma_\mu,\gamma_\nu\} = 2\delta_{\mu\nu}$; $\gamma_\mu^\dagger = \gamma_\mu$; $\gamma_5= \gamma_4\gamma_1\gamma_2\gamma_3$, tr$[\gamma_5\gamma_\mu\gamma_\nu\gamma_\rho\gamma_\sigma]=-4 \epsilon_{\mu\nu\rho\sigma}$; $\sigma_{\mu\nu}=(i/2)[\gamma_\mu,\gamma_\nu]$; $a \cdot b = \sum_{i=1}^4 a_i b_i$; and $P_\mu$ timelike $\Rightarrow$ $P^2<0$.}
$n_\mu$ is a lightlike four-vector, $n^2=0$, $z\cdot P = x\, n\cdot P$, $n\cdot P = -m_\rho$,
$\epsilon_\mu^\lambda$ is a polarisation four-vector;
$f_\rho$ and $f_\rho^\perp$ are, respectively, $\rho$-meson vector and tensor decay constants, with the former, a renormalisation point invariant, explaining the strength of $\rho \to e^+ e^-$ decay; and $\zeta$ is the renormalisation scale.  The tensor decay constant depends on $\zeta$.  With these conventions,
\begin{equation}
\int_0^1 dx\, \varphi_\|(x,\zeta) = 1 = \int_0^1 dx\, \varphi_\bot(x,\zeta)\,.
\end{equation}

Note that in order to produce quantities that are gauge invariant for all values of $z$, each of the left-hand-sides in Eq.\,\eqref{rhoPDAs} should also contain a Wilson line:
\begin{equation}
{\cal W}[-z/2,z/2] = \exp ig\int_{-z/2}^{z/2} d \sigma_\mu A_\mu(\sigma)\,,
\end{equation}
between the quark fields.  Plainly, for any light-front trajectory, ${\cal W}[-z/2,z/2]\equiv 1$ in lightcone gauge: $n\cdot A=0$, and hence the Wilson line does not contribute when this choice is employed.  On the other hand, light-cone gauge is seldom practicable in either model calculations or quantitative nonperturbative analyses in continuum QCD.  Herein, indeed, as is typical in nonperturbative DSE studies, we employ Landau gauge because, \emph{inter alia} \cite{Bashir:2008fk,Bashir:2009fv,Raya:2013ina}: it is a fixed point of the renormalisation group; that gauge for which sensitivity to model-dependent differences between \emph{Ans\"atze} for the fermion--gauge-boson vertex are least noticeable; and a covariant gauge, which is readily implemented in numerical simulations of lattice-regularised QCD.  It is therefore significant that ${\cal W}[-z/2,z/2]$ is not quantitatively important in the calculation of the leading-twist contributions to matrix elements like those in Eqs.\,\eqref{rhoPDAs} \cite{Kopeliovich:2011rv}.

It is evident from Eqs.\,\eqref{rhoPDAs} that four leading-twist PDAs are associated with the $\rho$-meson: $\varphi_\|(x)$, $\varphi_\perp(x)$ describe, respectively, the light-front fraction of the $\rho$-meson's total momentum carried by the quark in a longitudinally or transversely polarised $\rho$; and $g^v_\perp(x)$, $g^a_\perp(x)$ are analogous quantities associated with transversely polarised quarks in a longitudinally polarised $\rho$-meson.  However, only two of these four amplitudes are independent at leading twist \cite{Ball:1998sk}; viz., with $\bar{v}=1-v$ and $\bar{x}=1-x$,
\begin{eqnarray}
g_\bot^{(v)}(x,\zeta) &=&
\frac{1}{2}\left[\int^x_0 \! dv\frac{\varphi_\|(v,\zeta)}{\bar{v}}
+\int^1_x \!  dv\frac{\varphi_\|(v,\zeta)}{v}\right],\\
g_\bot^{(a)}(x,\zeta) &=&
2\left[\bar{x}\int^x_0 \!  dv\frac{\varphi_\|(v,\zeta)}{\bar{v}}
+ x \int^1_x \!  dv\frac{\varphi_\|(v,\zeta)}{v}\right].\quad
\end{eqnarray}

The $\zeta$-dependence of the PDAs is important: it specifies the mass-scale relevant to the process in which the meson is involved and hence at which the PDA is to be employed; and the shape of a given PDA changes with $\zeta$.  QCD is invariant under the collinear conformal group on the domain $\tau \Lambda_{\rm QCD} \simeq 0$  \cite{Brodsky:1980ny,Braun:2003rp}, where we have defined $\tau=1/\zeta$.  It follows that twist-two pseudoscalar and vector meson PDAs are accurately approximated on that domain by
\begin{eqnarray}
\label{PDAG3on2}
\varphi(x;\tau) &=& \varphi^{\rm asy}(x)
\bigg[ 1 + \!\! \sum_{j=2,4,\ldots}^{\infty} \!\! \!\! a_j^{3/2}(\tau) \,C_j^{(3/2)}(2 x -1) \bigg],\;\;\\
\varphi^{\rm asy}(x) &=& 6 x (1-x)\,, \label{phiasy}
\end{eqnarray}
where $\{C_j^{(3/2)}|j=1,\ldots,\infty\}$ are Gegenbauer polynomials of order $\alpha=3/2$ and the expansion coefficients $\{a_j^{3/2}|j=1,\ldots,\infty\}$ evolve logarithmically with $\tau$, vanishing as $\tau\Lambda_{\rm QCD}\to 0$.

Until recently it was commonly assumed that at any length-scale, $\tau$, an accurate approximation to $\varphi(x;\tau)$ is obtained by using just the first few terms of the expansion in Eq.\,\eqref{PDAG3on2}.  Let us call this \emph{Assumption~A}.  It has led to models for $\varphi(x;\tau)$ whose pointwise behaviour is not concave on $x\in[0,1]$; e.g., to ``humped'' distributions \cite{Chernyak:1983ej}.  Following Ref.\,\cite{Chang:2013pq}, one may readily establish that a double-humped form for $\varphi(x)$ lies within the class of distributions produced by a meson Bethe-Salpeter amplitude which may be characterised as vanishing at zero relative momentum, instead of peaking thereat.  No ground-state pseudoscalar or vector meson Bethe-Salpeter equation solution exhibits corresponding behaviour \cite{Maris:1997tm,Maris:1999nt,Qin:2011xq}.

\emph{Assumption~A} is certainly valid on $\tau \Lambda_{\rm QCD} \simeq 0$.  However, as we shall illustrate again herein, it is unsound at any energy scale accessible in contemporary or foreseeable experiments.  This was highlighted in Ref.\,\cite{Cloet:2013tta} and in Sec.\,5.3 of Ref.\,\cite{Cloet:2013jya}.  The latter used the fact \cite{Georgi:1951sr,Gross:1974cs,Politzer:1974fr} that $\varphi^{\rm asy}(x)$ can only be a good approximation to a meson's PDA when it is accurate to write $u_{\rm v}(x) \approx \delta(x)$, where $u_{\rm v}(x)$ is the meson's valence-quark PDF, and showed that this is not valid even at energy scales characteristic of the large hadron collider (LHC).  An identical conclusion was reached in Ref.\,\cite{Segovia:2013eca}, via consideration of the first moment of the kaon's valence-quark PDA, which is a direct measure of SU$(3)$-flavour breaking and must therefore vanish in the conformal limit.  Hence, realistic meson PDAs are necessarily broader than $\varphi^{\rm asy}(x)$.  It follows that an insistence on using just a few terms in Eq.\,\eqref{PDAG3on2} to represent a hadron's PDA will typically lead to unphysical oscillations; i.e., humps, just as any attempt to represent a box-like curve via a Fourier series will inevitably lead to slow convergence and spurious oscillations.  In our analysis of vector meson PDAs, we will use an alternative to \emph{Assumption~A}, which is explained in the next subsection.

\subsection{Computing light-front projections of \mbox{\boldmath $\psi_{\rm BS}$}}
\label{ComputingPsiBS}
Returning now to Eqs.\,\eqref{rhoPDAs}, and making use of the properties of polarisation vectors and the relationship between Bethe-Salpeter wave functions in configuration and momentum space, one obtains
\begin{subequations}
\label{momPDAs}
\begin{eqnarray}
\nonumber \lefteqn{f_\rho \,n\cdot P\, \varphi_\|(x;\zeta) =
m_\rho {\rm tr}_{\rm CD} Z_2(\zeta,\Lambda)}\\
&& \times \int_{dq}^\Lambda
\delta(n\cdot q_+ - x n\cdot P) \gamma\cdot n \,n_\nu \chi_\nu(q;P)\,,
\label{frho}\\
\nonumber \lefteqn{f_\rho^\perp \,m_\rho^2 \, \varphi_\perp(x;\zeta) =
 n\cdot P\,  {\rm tr}_{\rm CD} Z_T(\zeta,\Lambda)}\\
&& \times \int_{dq}^\Lambda
\delta(n\cdot q_+ - x n\cdot P) \sigma_{\mu\nu} P_\mu \chi_\nu(q;P)\,,
\label{frhoT}
\end{eqnarray}
\end{subequations}
where: $\int_{dq}^\Lambda $ is a Poincar\'e-invariant regularisation of the four-dimensional integral, with $\Lambda$ the ultraviolet regularisation mass-scale; and $Z_{2,T}(\zeta,\Lambda)$ are, respectively, the renormalisation constants for the quark wave-function and the tensor vertex.  (See Appendix~\ref{Z2ZT}.)

In Eqs.\,\eqref{momPDAs} the Bethe-Salpeter wave function is
\begin{equation}
\label{BSwavefunction}
\chi_\nu(q;P) = S(q_+) \Gamma^\rho_\nu(q;P) S(q_-)\,,
\end{equation}
where: $S$ is the dressed propagator, which takes the form in Eq.\,\eqref{SpDressed}; $q_+=q+\eta P$, $q_-=q-(1-\eta)P$, $\eta\in[0,1]$; and
\begin{eqnarray}
\label{rhoBSA}
\Gamma_\nu(q;P)=\sum_{j=1}^8 \tau_\nu^j(q,P) F^j(q^2,q\cdot P;P^2)
\end{eqnarray}
is the $\rho$-meson Bethe-Salpeter amplitude, with the Dirac-matrix tensor basis $\{\tau_\nu^j|j=1,\ldots,8\}$ defined in Eq.\,\eqref{eq:bsa}.  Owing to Poincar\'e invariance, no observable can legitimately depend on $\eta$; i.e., the definition of the relative momentum.  On the other hand, the choice $\eta=1/2$ is computationally convenient when working with a multiplet that contains a charge-conjugation eigenstate because then the scalar functions $\{F^j(q^2,q\cdot P;P^2)|j=1,\ldots,8\}$ in Eq.\,\eqref{rhoBSA} are even under $q\cdot P \to (-q\cdot P)$.

With $\chi_\nu(q;P)$ in hand, it is straightforward to follow the procedure explained in Ref.\,\cite{Chang:2013pq} and thereby obtain $\varphi_{\|,\perp}$ from Eqs.\,\eqref{momPDAs}.  The first step is to compute the moments
\begin{subequations}
\label{momentsE}
\begin{eqnarray}
\nonumber
\lefteqn{\langle x^m \rangle_\| = \int_0^1 dx \, x^m\, \varphi_\|(x)}\\
&=&
\frac{m_\rho}{f_\rho}{\rm tr}_{\rm CD} Z_2
\int_{dq}^\Lambda \frac{[n\cdot q_+]^m}{[n\cdot P]^{m+2}}\,
\gamma\cdot n \,n_\nu \chi_\nu(q;P)\,,\quad
\label{momentsParellel}\\
\nonumber
\lefteqn{\langle x^m \rangle_\perp = \int_0^1 dx \, x^m\, \varphi_\perp(x)}\\
&=& \frac{1}{f_\rho^\perp  m_\rho^2 }  {\rm tr}_{\rm CD} Z_T
\int_{dq}^\Lambda
\frac{[n\cdot q_+]^m}{[n\cdot P]^{m}}\, \sigma_{\mu\nu} P_\mu \chi_\nu(q;P)\,.\quad
\end{eqnarray}
\end{subequations}
Notably, beginning with an accurate form of $\chi_\nu(q;P)$, arbitrarily many moments can be computed.

One then capitalises on the fact that Gegenbauer polynomials of order $\alpha$, $\{C_n^{\alpha}(2 x -1)| n=0,\ldots,\infty\}$, are a complete orthonormal set on $x\in[0,1]$ with respect to the measure $[x (1-x)]^{\alpha_-}$, $\alpha_-=\alpha-1/2$, and hence they enable reconstruction of any function defined on $x\in[0,1]$ that vanishes at the endpoints. (N.B.\, Owing to charge-conjugation invariance, $\varphi_{\|,\perp}(x)$ are even under $x\leftrightarrow \bar x$; and they vanish at the endpoints unless the underlying interaction is momentum-independent.)
Therefore, with complete generality, the PDA for any member of a multiplet containing an eigenstate of charge conjugation may accurately be approximated as follows:
\begin{equation}
\label{PDAGalpha}
\varphi(x) \approx  \varphi_{m}(x) =N_\alpha [x \bar x]^{\alpha_-}
\bigg[ 1 + \sum_{j=2,4,\ldots}^{j_{\rm max}} a_j^\alpha C_j^\alpha(x -\bar x) \bigg],
\end{equation}
where $N_\alpha = \Gamma(2\alpha+1)/[\Gamma(\alpha+1/2)]^2$.

Finally, from a given set of $m_{\rm max}$ moments computed via Eqs.\,\eqref{momentsE}, the PDA is determined by minimising
\begin{equation}
\varepsilon_m = \sum_{l=2,4,\ldots,m_{\rm max}} |\langle x^l\rangle_{m}/\langle x^l\rangle-1|\,,
\end{equation}
over the set $\{ \alpha, a_2, a_4, \ldots, a_{j_{\rm max}}\}$, where
\begin{equation}
\label{endprocedure}
\langle x^l\rangle_{m} = \int_0^1 dx\, x^l \varphi_{m}(x)\,.
\end{equation}

This is the alternative to \emph{Assumption A}, mentioned above and exploited elsewhere \cite{Chang:2013pq,Cloet:2013tta,Chang:2013epa,Cloet:2013jya,Segovia:2013eca}.  It recognises that, at all accessible scales, the pointwise profile of PDAs is determined by nonperturbative dynamics; and hence PDAs should be reconstructed from moments by using Gegenbauer polynomials of order $\alpha$, with this order -- the value of $\alpha$ -- determined by the moments themselves, not fixed beforehand.  In the pion case, this procedure converged very rapidly: $j_{\rm max}=2$ was sufficient \cite{Chang:2013pq}.

Naturally, once obtained in this way, one may project $\varphi(x;\tau)$ onto the form in Eq.\,\eqref{PDAG3on2}; viz., for $j=1,2,\ldots\,$,
\begin{equation}
\label{projection}
a_j^{3/2}(\tau) = \frac{2}{3}\ \frac{2\,j+3}{(j+2)\,(j+1)}\int_0^1 \!dx\, C_j^{(3/2)}(x-\bar x)\,\varphi(x;\tau),
\end{equation}
therewith obtaining all coefficients necessary to represent any computed distribution in the conformal form without ambiguity or difficulty.  It is then straightforward to determine the distribution at any $\tau^\prime < \tau$ using the appropriate evolution equations for the coefficients $\{a_j^{3/2}(\tau),i=1,2,\ldots\}$ \cite{Efremov:1979qk,Lepage:1980fj,Ball:1996tb,Ball:1998sk,Braun:2003rp}.

\section{Results}
\label{secResults}
\subsection{Algebraic example}
In order to reliably compute moments via Eqs.\,\eqref{momentsE}, we follow Ref.\,\cite{Chang:2013pq} and develop a Nakanishi-like representation \cite{Nakanishi:1963zz,Nakanishi:1969ph,Nakanishi:1971} of the vector meson Bethe-Salpeter wave functions that are ultimately obtained via numerical solution of a Bethe-Salpeter equation.  Before detailing the results of such numerical analysis, we judge it useful to provide a simple algebraic illustration of this idea.  Therefore, consider Eq.\,\eqref{momentsParellel} and write
\begin{subequations}
\label{NakanishiASY}
\begin{eqnarray}
\label{eq:sim1}
S(p)&=&[-i\gamma\cdot p+M]\Delta_M(p^2)\,,\\
\rho_\nu(z)&=&\frac{1}{\sqrt{\pi}}\frac{\Gamma(v+3/2)}{\Gamma(\nu+1)}(1-z^2)^\nu\,,\\
\label{eq:sim2}
\Gamma_\mu(q;P)&=&i\gamma_\mu^{T}\frac{M}{f_\rho}\int^1_{-1}dz\rho_\nu(z) \hat\Delta^\nu_M(q^2_{+z})\,,
\end{eqnarray}
\end{subequations}
where $\Delta_M(s)=1/[s+M^2]$, $\hat\Delta_M(s)=M^2\Delta_M(s)$, $q_{+z}=q-(1-z)P/2$.  In the resulting expression, using a Feynman parametrisation, the three denominators can be combined into one $q$-quadratic form, raised to a power that depends linearly on $\nu$.  A subsequent change of variables enables one to isolate the $d^4q$ integration and arrive at
\begin{eqnarray}
\nonumber
\langle x^m \rangle_{\varphi_\|} &=& {\rm constant}\int^1_0 dx \, x^{\nu-1} \int^{1-x}_{0}dy \int^1_{-1} dz \rho_{\nu}(z)u^m\\
&& \times {\mathpzc D}^{-\nu}\,   [{\mathpzc D}+(\nu-1)(M^2-u(1-u)P^2)]\,,
\label{momUV}
\end{eqnarray}
where ``constant'' involves the momentum integral; $x$, $y$ are Feynman parameters; $u=\frac{x(1+z)}{2}+y$; and, with $P^2=-m_\rho^2$,
\begin{equation}
{\mathpzc D}=M^2+[\mbox{$\frac{1}{4}$}x(1-x)(1+z)^2+y(1-y)-x y(1+z)]P^2.
\end{equation}

In a QCD-like theory one has $\nu=1$ and the Bethe-Salpeter amplitude behaves as $1/q^2$ in the ultraviolet.  In this case, capitalising on the connection that exists for any $\nu$ between ``constant'' and the canonical PDA normalisation, Eq.\,\eqref{momUV} collapses to
\begin{eqnarray}
\nonumber
\langle x^m \rangle_{\varphi_\|} &=& \int^1_0 dx \, x^{\nu-1} \int^{1-x}_{0}dy \int^1_{-1} dz \rho_{\nu}(z)u^m \\
&=& \frac{6}{(m+2) (m+3)}\,,
\label{momUV1}
\end{eqnarray}
from which one immediately obtains
\begin{equation}
\label{phiisphiasy}
\varphi_\|(x) = \varphi^{\rm asy}(x)\,;
\end{equation}
namely, the spectral representation of the Bethe-Salpeter wave function produced by Eqs.\,\eqref{NakanishiASY} yields the asymptotic PDA.  A similar analysis produces the same result for $\varphi_\perp(x)$.  In fact, for both pseudoscalar- and vector-mesons, this procedure produces the correct asymptotic distribution for the leading twist contribution to any given projection of the Bethe-Salpeter wave function in a  $(1/k^2)^\nu$ vector exchange theory \cite{Chang:2013pq,Chang:2013epa}.

\subsection{Meson properties and PDAs}
\subsubsection{Rainbow-ladder truncation}
We turn now to our detailed analysis of vector meson valence-quark distribution amplitudes, which employs gap and Bethe-Salpeter equation solutions obtained using the rainbow-ladder truncation of QCD's DSEs \cite{Maris:2003vk,Chang:2011vu,Bashir:2012fs,Cloet:2013jya} and the interaction introduced in Ref.\,\cite{Qin:2011dd}.
Pertinent details are presented in Appendix~\ref{sec:kernels}.
Here we simply observe that the rainbow-ladder truncation is the leading order in a systematic, symmetry preserving procedure that enables a tractable formulation of the continuum bound-state problem \cite{Munczek:1994zz,Bender:1996bb}.  It is a widely used DSE computational scheme in hadron physics and is known to be accurate for ground-state vector- and isospin-nonzero-pseudoscalar-mesons \cite{Maris:2003vk,Chang:2011vu,Bashir:2012fs,Cloet:2013jya}, and properties of the nucleon and $\Delta$-resonance \cite{Eichmann:2011ej,Chen:2012qr,Segovia:2013rca,Segovia:2013uga}, because corrections in these channels largely cancel owing to parameter-free preservation of the relevant Ward-Takahashi identities.
Concerning the interaction in Ref.\,\cite{Qin:2011dd}, its infrared composition is deliberately consistent with that determined in modern studies of QCD's gauge sector, which indicate that the gluon propagator is a bounded, regular function of spacelike momenta, $q^2$, that achieves its maximum value on this domain at $q^2=0$ \cite{Bowman:2004jm,Cucchieri:2011ig,Boucaud:2011ug,Ayala:2012pb,Aguilar:2012rz,%
Strauss:2012dg}, and the dressed-quark-gluon vertex does not possess any structure which can qualitatively alter these features \cite{Skullerud:2003qu,Bhagwat:2004kj}.  It also preserves the one-loop renormalisation group behaviour of QCD so that, e.g., the quark mass-function is independent of the renormalisation point.
We list our calculated values for vector meson static properties in Table~\ref{tablestatic}.

\begin{table}[t]
\caption{Vector meson static properties computed using RL truncation with the interaction in Eq.\,\eqref{CalGQC} and renormalisation point invariant current-quark masses $\hat m_{u/d}=6.4\,$MeV, $\hat m_s = 143\,$MeV, which correspond to the following one-loop evolved values $m_{u/d}^{\zeta=2\,{\rm GeV}} \stackrel{\rm 1\,loop}{=}4.5\,$MeV, $m_{s}^{\zeta=2\,{\rm GeV}} \stackrel{\rm 1\,loop}{=}99\,$MeV.
(All quantities tabulated in GeV.)
For comparison, we also list selected results obtained elsewhere, using:
models of light-cone wave functions \cite{Forshaw:2003ki};
QCD sum rules \cite{Ball:1996tb,Ball:1998sk};
nonlocal condensates \cite{Bakulev:1998pf,Pimikov:2013usa};
light-front quantum mechanics \cite{Choi:2007yu};
and AdS/QCD models for light cone wave functions \cite{Ahmady:2012dy}.  In those studies, empirical masses were used as input constraints.
The last two rows report results from lattice-QCD \cite{Braun:2003jg,Jansen:2009hr}.
%
Some experimental values are \protect\cite{Beringer:1900zz}:
$m_{u/d}^{\zeta=2\,{\rm GeV}}=3.6^{+0.6}_{-0.4}\,$MeV;
$m_{s}^{\zeta=2\,{\rm GeV}}=95 \pm 5\,$MeV;
$f_\rho=0.153\,$GeV, $m_\rho=0.777\,$GeV; and $f_\phi=0.168\,$GeV, $m_\phi=1.02\,$GeV.
\label{tablestatic}
}
\begin{tabular*}
{\hsize}
{
l|@{\extracolsep{0ptplus1fil}}
l|@{\extracolsep{0ptplus1fil}}
l|@{\extracolsep{0ptplus1fil}}
l|@{\extracolsep{0ptplus1fil}}
l|@{\extracolsep{0ptplus1fil}}
l|@{\extracolsep{0ptplus1fil}}
l@{\extracolsep{0ptplus1fil}}}\hline
      & $m_\rho$ & $m_\phi$  & $f_\rho$ & $f_\rho^\perp$ & $f_\phi$ & $f_\phi^\perp$ \\\hline
herein & $0.74$ & $1.08$ & $0.15$ & $0.11$ & $0.19$ & $0.15$\\\hline
%
\protect\cite{Forshaw:2003ki} 
        &   &   & $0.17(4)$ & $0.15(1)$ & $0.084(12)$ & $0.078(2)$\\
\protect\cite{Ball:1996tb,Ball:1998sk} 
        &   &   &   & $0.11(1)$ &  & $0.15(1)$\\
\protect\cite{Bakulev:1998pf}  
        &   &   & $0.14(1)$  & $0.12(1)$ & & \\
\protect\cite{Pimikov:2013usa} 
        &   &   & $0.15(1)$  & $0.12(1)$ & & \\
\protect\cite{Choi:2007yu}  
        &   &   & $0.16(2)$ & $0.13(1)$ & & \\
\protect\cite{Ahmady:2012dy} 
        &   &   & $0.15$ & $0.095$ & & \\\hline
\protect\cite{Braun:2003jg} 
        & 0.79(4)  & 1.00(1)  &  & $0.113(2)$ & & 0.131(1) \\
\protect\cite{Jansen:2009hr} 
        & 0.90(4) & 1.13(8)  & $0.17(1)$ & $0.13(1)$ & $0.22(2)$ & $0.16(2)$ \\\hline
\end{tabular*}
\end{table}

\subsubsection{DCSB in vector mesons}
The values and ratios of the vector meson decay constants reveal some interesting features of these mesons.  They can be exposed by noting from Eqs.\,\eqref{rhoPDAs} that $f_{\rho,\phi}$ are associated with currents which are invariant under chiral transformations.  On the other hand, the currents that define $f_{\rho,\phi}^\perp$ are not chirally invariant and hence the values of $f_{\rho,\phi}^\perp$ are a clear expression of the strength of chiral symmetry breaking within the vector mesons.  At $\zeta=2\,$GeV,
\begin{equation}
\label{rhoDCSB}
f^\perp_\rho/f_\rho = 0.73\,,\; f^\perp_\phi/f_\phi = 0.79\,;
\end{equation}
and thus, at an hadronic scale, chiral symmetry breaking is strong within the $\rho$- and $\phi$-mesons.  Since the scale of explicit chiral symmetry breaking is small for light quarks, it is DCSB that is expressed prominently in the large values of $f_{\rho,\phi}^\perp$.  (This is analogous to the connection between DCSB and the appearance of dressed-quark anomalous chromo- and electro-magnetic moments \cite{Singh:1985sg,Kochelev:1996pv,Bicudo:1998qb,Chang:2010hb,Bashir:2011dp,Qin:2013mta}.)

Naturally, DCSB is also expressed in the magnitudes of $f_{\rho,\phi}$.  This situation is kindred to the relationship between $f_\pi$ and the chiral condensate, $\rho_\pi$ \cite{Brodsky:2008be,Brodsky:2009zd,Brodsky:2010xf,Chang:2011mu,Brodsky:2012ku,Cloet:2013jya}, both of which are order parameters for DCSB.  There are differences, too, however, because the matrix element connected with $\rho_\pi$ receives its first contribution at twist-three, the associated distribution function is almost frozen under evolution \cite{Chang:2013epa} and, indeed, $\rho_\pi$ actually increases with increasing $\zeta$, whereas the ratios $f^\perp_\rho/f_\rho$, $f^\perp_\phi/f_\phi$ evolve to zero with increasing $\zeta$ (see Appendix~\ref{Z2ZT}).

The chiral transformation properties of the currents in Eqs.\,\eqref{rhoPDAs} are independent of the parton content of the vector mesons and the decay constants $f_{\rho,\phi}$, $f_{\rho,\phi}^T$ are Poincar\'e invariant, so the statements just made are independent of the reference frame.

Suppose now that one assumes a vector meson is an instant-form \cite{Dirac:1949cp} quantum mechanical bound-state of a dressed-quark and -antiquark, which themselves are vectors in the Hilbert space associated with some well-defined instant-form Hamiltonian.
Under these circumstances, it has been argued \cite{Glozman:2011nk} that it is possible to draw a connection between the ratios $f^\perp_\rho/f_\rho$, $f^\perp_\phi/f_\phi$ and the angular momentum configurations in the centre-of-momentum frame of those same dressed quarks:
\begin{equation}
\label{SDratio1}
\left(\begin{array}{c}
a_{\,^3\!S_1}\\
a_{\,^3\!D_1}
\end{array}\right)
=
\left(\begin{array}{cr}
\sqrt{\frac{2}{3}} & \sqrt{\frac{1}{3}} \\
\sqrt{\frac{1}{3}} & -\sqrt{\frac{2}{3}}
\end{array}\right)
\left(\begin{array}{c}
\frac{f}{\sqrt{f^2 + (f^T)^2}}\\
\frac{f^T}{\sqrt{f^2 + (f^T)^2}}
\end{array}\right)\,,
\end{equation}
where $a_{\,^3\!S_1}$, $a_{\,^3\!D_1}$ are probability amplitudes describing the $S$- and $D$-state content of the vector meson.  In the present case, this would mean
\begin{equation}
\label{SDratio2}
\left(\begin{array}{ccr}
a^\rho_{\,^3\!S_1}&=&1.00\\
a^\rho_{\,^3\!D_1}&=&-0.02
\end{array}\right)\,,
\left(\begin{array}{ccr}
a^\phi_{\,^3\!S_1}&=&1.00\\
a^\phi_{\,^3\!D_1}&=&-0.05
\end{array}\right)\,;
\end{equation}
i.e., the vector mesons are almost purely $S$-wave states at $\zeta=2\,$GeV.  Given the evolution of $f^T$ with $\tau=1/\zeta$, one would also be led to conclude that the $S/D$-wave ratio reaches a maximum value of $\sqrt{2}$ on $\tau \Lambda_{QCD} \simeq 0$.

Prompted by these assertions, which seem unlikely outcomes in the Poincar\'e covariant treatment of a light-quark bound-state, we considered the structure of our computed Bethe-Salpeter wave function; namely, $\chi_\nu(q;P)$ in Eq.\,\eqref{BSwavefunction}.  When expressed in terms of the complete orthogonal tensor basis in Eq.\,\eqref{eq:bsa}:
\begin{equation}
\label{eq:chi}
\chi_\nu(q;P) = \sum_{j=1}^8 \tau_\nu^j(q,P) \,{\mathpzc F}_\chi^j(q^2,q\cdot P;P^2)\,,
\end{equation}
we find that with $\vec{P}=0$ only ${\mathpzc F}_\chi^2$ might be considered small, $\{{\mathpzc F}_\chi^{j=4,5,7,8}\}$ are non-negligible and commensurate in magnitude, and $\{{\mathpzc F}_\chi^{j=3,6}\}$ are $3$-$4$-times larger than those functions in the infrared.  Notably, $f^T$ is a direct measure of $[4 {\mathpzc F}_\chi^7 + {\mathpzc F}_\chi^8]$: no other function in Eq.\,\eqref{eq:chi} contributes.

It is straightforward to translate these observations into statements about the scalar functions that appear when the basis in Ref.\,\cite{LlewellynSmith:1969az} is used.  That basis has the merit that it allows one to readily identify those functions which correspond to $\,^3\!S_1$ and $\,^3\!D_1$ components within a vector meson at rest: $S$-wave, $A,B,C,D$; and $D$-wave, $E,F,G,H$.  Using our computed solutions, we find that $B,C,G$ are large, $A,D,E$ are material and of similar magnitude, and $F,H$ are small.  The sizes of $E,G$ indicate that the $\,^3\!D_1$ components are significant.  On the other hand, since $H$ is small, $f^T$ receives its largest contribution from $[2 B (P\cdot q)^2 + C P^2]$; viz., from $S$-wave components.

A logical way to quantify the size of a vector meson's $D$-state is to compute the contribution from $E,F,G,H$ to the Bethe-Salpeter wave function's Poincar\'e-invariant canonical normalisation constant, ${\mathpzc N}^2$ \cite{Bhagwat:2006xi,Cloet:2007pi}.  The algebraic correspondence between the bases entails that
\begin{equation}
\label{Dequal0}
\,^3\!D_1\,\mbox{component} = 0 \; \mbox{when} \; {\mathpzc F}_\chi^{2,3,5} = 0 \,,\;
{\mathpzc F}_\chi^{8} = 2\,{\mathpzc F}_\chi^{7}\,.
\end{equation}
Using this fact, our vector meson solutions yield
\begin{equation}
\label{ratioN2}
\frac{{\mathpzc N}^2_{\;\rho_{S+0}}}{{\mathpzc N}^2_{\;\rho_{S+D}}} = 0.59\,,\quad
\frac{{\mathpzc N}^2_{\;\phi_{S+0}}}{{\mathpzc N}^2_{\;\phi_{S+D}}} = 0.71\,.
\end{equation}

One may translate Eqs.\,\eqref{ratioN2} into the following expressions
\begin{subequations}
\label{SDratio2Real}
\begin{eqnarray}
&&\left(\begin{array}{ccrcr}
\tilde a^\rho_{\,^3\!S_1}&=& 0.77 &\approx& \sqrt{10/17}\\
\tilde a^\rho_{\,^3\!D_1}&=&-0.64 &\approx& -\sqrt{7/17}
\end{array}\right)\,,\\
&&\left(\begin{array}{ccrcr}
\tilde a^\phi_{\,^3\!S_1}&=& 0.85 &\approx& \sqrt{5/7}\\
\tilde a^\phi_{\,^3\!D_1}&=&-0.53 &\approx& -\sqrt{2/7}
\end{array}\right)\,;
\end{eqnarray}
\end{subequations}
viz., simple formulae that display, in a manner reminiscent of a quantum mechanical state vector, the relative strengths of $S$- and $D$-waves in composing the canonical normalisation of the Bethe-Salpeter amplitude.\footnote{Canonical normalisation of a Bethe-Salpeter wave function is a more complicated procedure than normalising a square-integrable wave function in quantum mechanics.  The expression is not diagonal in the squares of the scalar functions $\{{\mathpzc F}_\chi^j|j=1,\ldots,8\}$.  Nevertheless, Eqs.\,\eqref{SDratio2Real} represent an unambiguous accounting of the relative contribution to ${\mathpzc N}^2$ from the terms that correspond to $S$ and $D$ waves in the vector meson's rest frame.}
%
Plainly, the $D$-wave contribution is comparable with that of the $S$-wave but, naturally, its strength diminishes as the scale associated with chiral symmetry breaking in the system increases.  For light-quarks, that scale owes largely to DCSB.
The ratios in Eq.\,\eqref{ratioN2} decrease by less-than 7\% when the renormalisation point is shifted from $\zeta=2\,$GeV to $\zeta=19\,$GeV and hence, by this measure, the relative importance of the $D$-state does not change rapidly as $f^T$ decreases.

At this point, using solutions of the Bethe-Salpeter equation which provide a good description of vector meson static properties, we have arrived directly at a set of results and conclusions that conflict markedly with those inferred from Eqs.\,\eqref{SDratio1}, \eqref{SDratio2}.
Indeed, we anticipated this result because Poincar\'e covariance itself requires the presence of all eight terms in Eq.\,\eqref{eq:chi} and it would be a peculiar interaction between light partons which imposed Eq.\,\eqref{Dequal0}, even approximately, simply because $\vec{P}=0$.
We therefore judge that Eqs.\,\eqref{SDratio1}, \eqref{SDratio2} cannot be associated with any property of a vector meson that is accessible via its Poincar\'e-covariant Bethe-Salpeter amplitude.

It should be observed that the assumptions made in order to arrive at Eqs.\,\eqref{SDratio1}, \eqref{SDratio2} and the associated conclusions are generally invalid in quantum field theory.  Whilst a Poincar\'e-covariant bound-state amplitude can be defined and computed in the vector meson's rest-frame -- a fact that we have just used -- QCD's dressed light-quarks cannot define vectors in the state space of an instant-form Hamiltonian.  A frame-independent Fock-space expansion of such a dressed parton is only possible on the light-front, whose features may be realised through computation in the infinite momentum frame.  This reference frame is also the only one in which an interpretation of hadron wave functions in terms of constituent-parton state-space probabilities is possible.  Compounding the interpretational difficulties further, in order to be accurate at an hadronic scale, the Fock space expansion of a dressed light-quark requires an enumerable infinity of increasingly complex partonic vectors.  In principle, therefore, no rigorous connection can exist between even a reliable estimation of a rest-frame wave function alone and measurable light-quark spin distributions.  Such a connection can only be drawn when a Poincar\'e-covariant wave function is used to compute Poincar\'e-invariant quantities, as, e.g., in Ref.\,\cite{Roberts:2013mja}, which shows that a picture of the proton based on SU$(6)$ spin-flavour wave functions is inaccurate.

\subsubsection{Integral representations of DSE solutions}
\begin{figure}[t]
\includegraphics[width=0.80\linewidth]{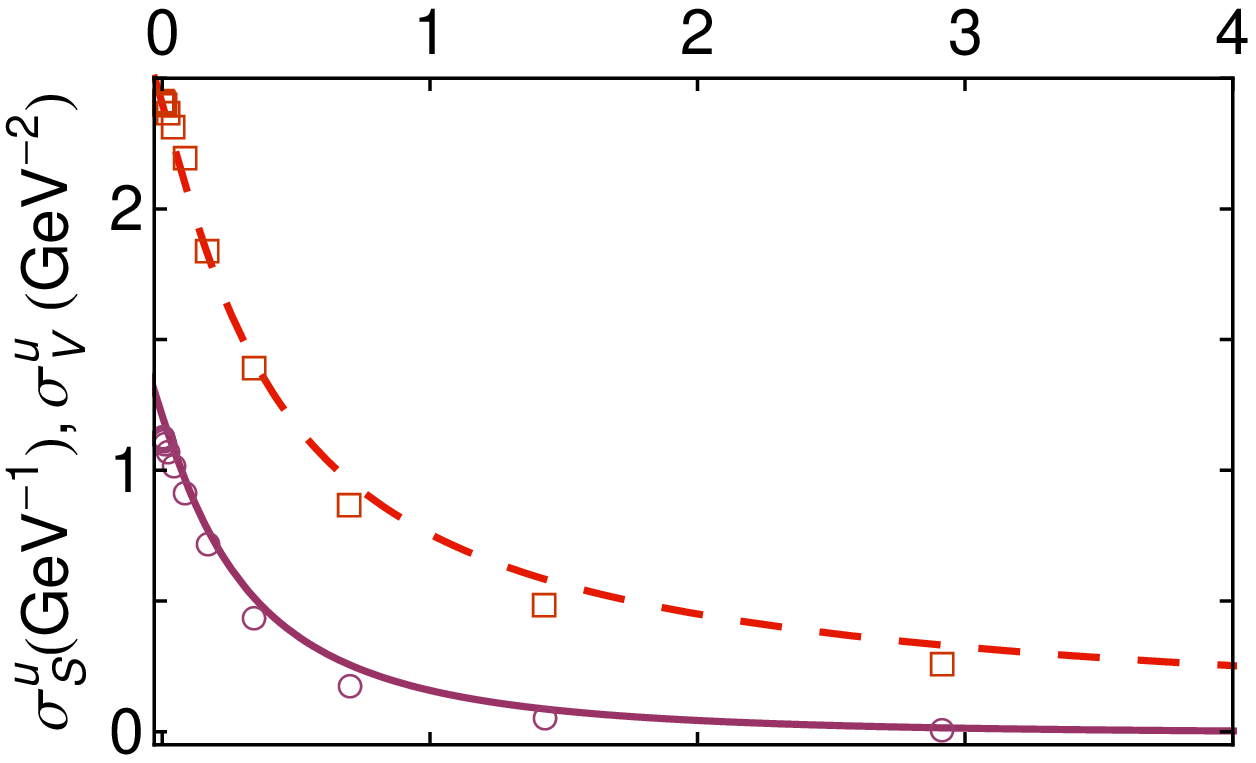}

\vspace*{-9.1ex}
\includegraphics[width=0.8\linewidth]{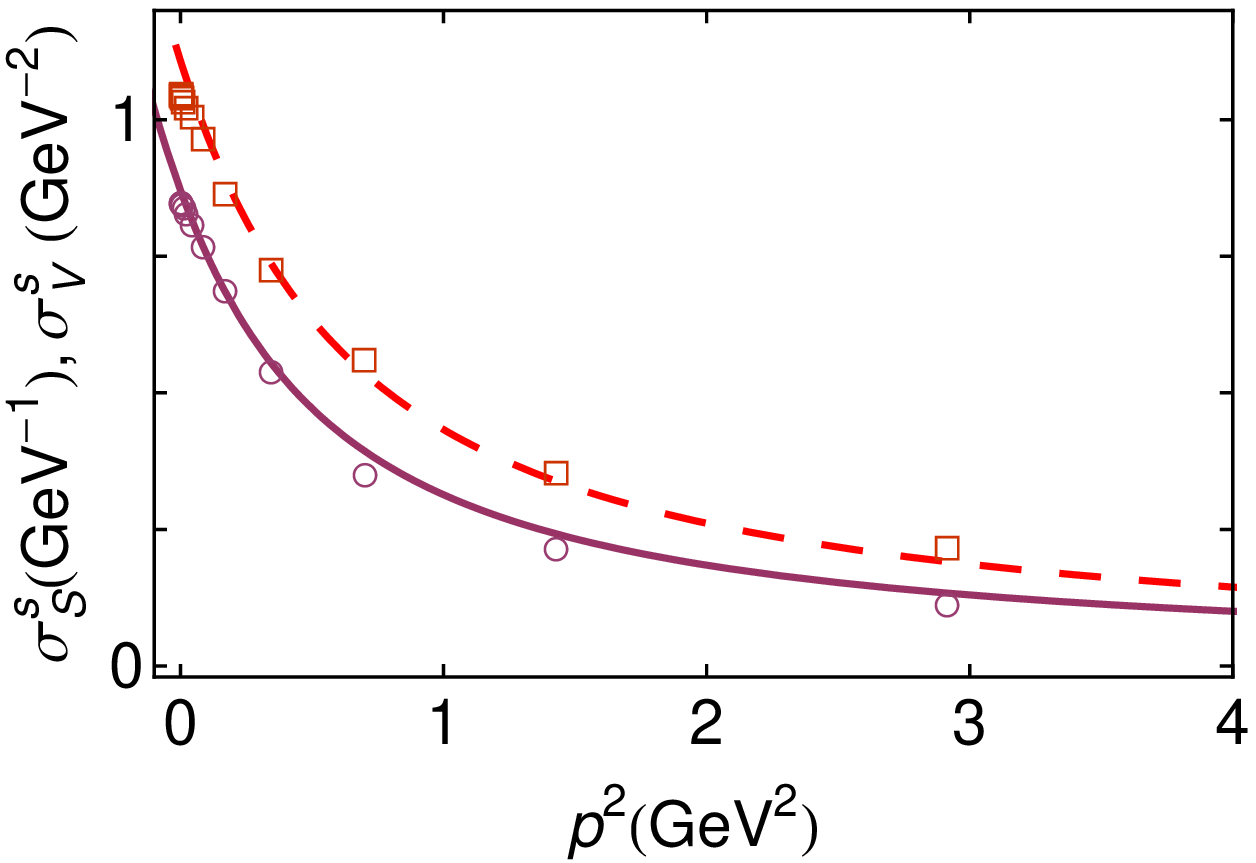}

\caption{Functions characterising the dressed quark propagator.
\emph{Upper panel}. $u/d$-quark functions, $\sigma_{S,V}^{u/d}(p^2)$ -- solution (open circles and squares, respectively) and interpolation functions (solid and long-dashed curves, respectively).
\emph{Lower panel}. $s$-quark\ functions, $\sigma_{S,V}^s(p^2)$, with same legend.
\label{fig:Splot}}
\end{figure}

The solutions of the gap and Bethe-Salpeter equations are obtained as matrices.  Computation of the moments in Eqs.\,\eqref{momentsE} is cumbersome with such input.  Indeed, brute numerical methods are completely inadequate to the task.  We therefore employ algebraic parametrisations of each array to serve as interpolations in evaluating the moments.  For the quark propagator, Eq.\,\eqref{SpDressed}, we represent $\sigma_{V,S}$ as meromorphic functions with no poles on the real $p^2$-axis \cite{Bhagwat:2002tx}, a feature consistent with confinement \cite{Roberts:2007ji,Chang:2011vu,Bashir:2012fs,Cloet:2013jya}.
%
Regarding the Bethe-Salpeter amplitudes, we retain all eight tensor structures and their associated functions because, even though the terms associated with $\{\tau_\nu^j|j=1,\ldots,5\}$ are largest in magnitude \cite{Maris:1999nt}, the $\{\tau_\nu^j|j=6,7,8\}$ structures are crucial to ensuring the correct magnitude and renormalisation group flow for the vector meson tensor decay constant, $f^T$.
Each of the functions is expressed via a Nakanishi-like representation \cite{Nakanishi:1963zz,Nakanishi:1969ph,Nakanishi:1971}; i.e., through integrals like Eq.\,\eqref{eq:sim2}, with parameters fitted to that function's first two nontrivial $q\cdot P$ Chebyshev moments.  (Details are presented in Appendix~\ref{NakanishiAppendix}.)  The quality of the description is illustrated via the dressed-quark propagators in Figs.\,\ref{fig:Splot}.

\subsubsection{Computed PDAs}
\begin{figure}[t]
\centerline{\includegraphics[width=0.5\textwidth]{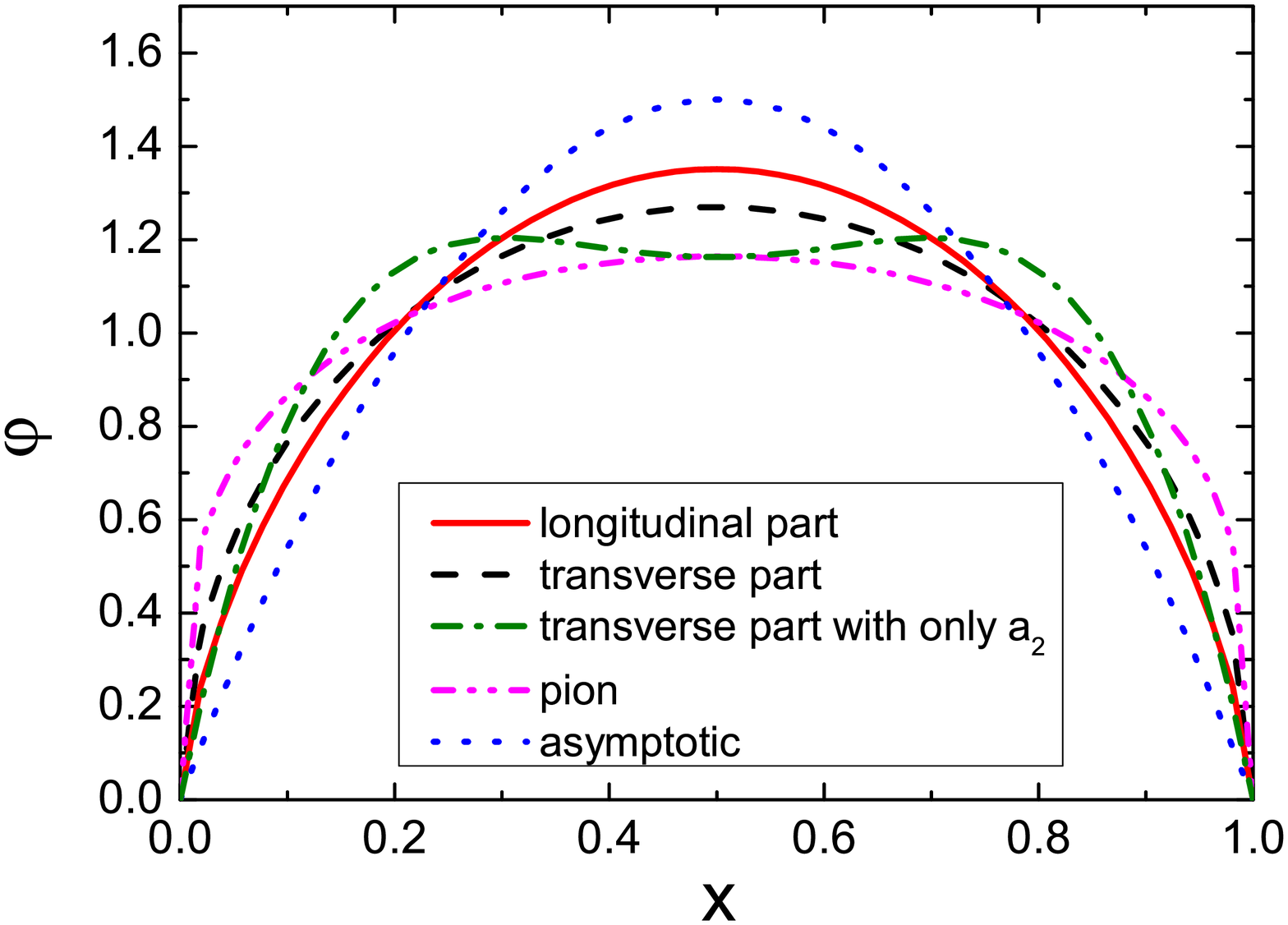}}
\caption{\label{fig:rda}  Computed distribution amplitudes for the $\rho$ meson. Curves:
\emph{solid}, $\varphi_\|(x)$;
\emph{dashed}, $\varphi_\perp(x)$;
\emph{dot-dashed}, $\varphi_\perp$ obtained using Eq.\eqref{PDAG3on2} and $a^{3/2}_{2\bot}=0.13$;
\emph{dot-dot-dashed}, RL result for $\phi_\pi(x)$, Eq.\,\eqref{resphipi2} \cite{Chang:2013pq};
and
\emph{dotted}, $\varphi^{\rm asy}(x)$ in Eq.\,\eqref{phiasy}.
}
\end{figure}

\begin{figure}[t]
\centerline{\includegraphics[width=0.5\textwidth]{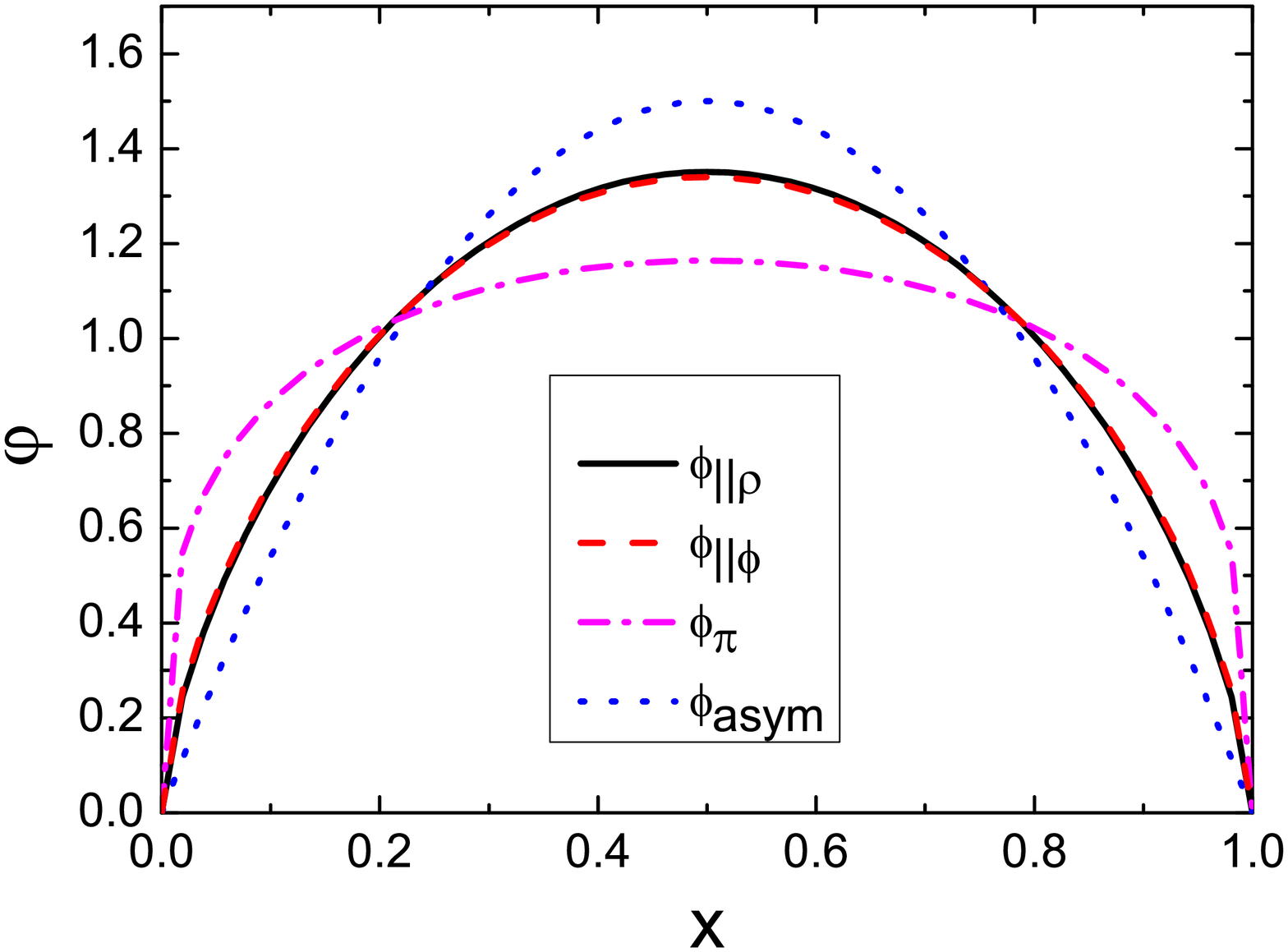}}
\centerline{\includegraphics[width=0.5\textwidth]{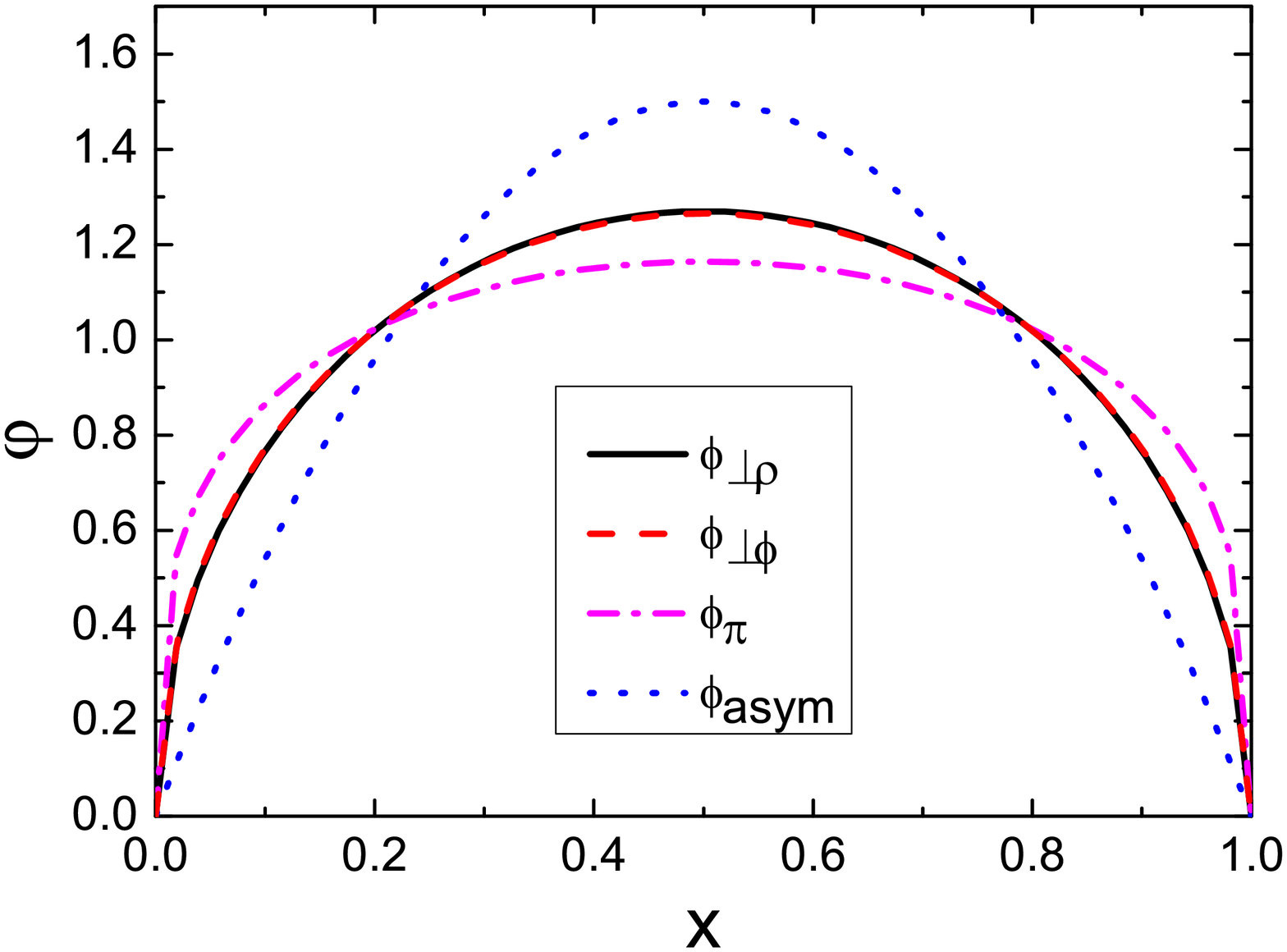}}
\caption{\label{fig:comp}
upper panel -- longitudinal distributions;
lower panel -- transverse distributions.
Curves: solid, $\rho$-meson; dashed, $\phi$-meson; dot-dashed, $\pi$-meson; and dotted, $\varphi^{\rm asy}$ in Eq.\,\eqref{phiasy}}
\end{figure}

Using the interpolating spectral representations, it is straightforward to compute arbitrarily many moments of the vector meson PDAs via Eqs.\,\eqref{momentsE}, following the pattern outlined in connection with Eqs,\,\eqref{NakanishiASY}--\eqref{phiisphiasy}.  We typically employ $m_{\rm max}=50$.  The pointwise forms of the PDAs are then reconstructed via the ``Gegenbauer-$\alpha$'' procedure described in connection with Eqs.\,\eqref{PDAGalpha}--\eqref{endprocedure} above.  The procedure converges very rapidly for the $\rho$- and $\phi$-mesons, so that, for both their longitudinal and transverse polarisations, the results obtained with $j_{\rm max}=2$ are indistinguishable, within line width, from those obtained when the sum in Eq.\,\eqref{PDAGalpha} is neglected completely.  Our results, depicted in Figs.\,\ref{fig:rda} and \ref{fig:comp}, are described by (renormalisation scale $\zeta_2=2\,$GeV):
\begin{subequations}
\label{phiresults}
\begin{eqnarray}
\phi_{\rho_\|}(x;\zeta_2) &=& 3.26\, x^{0.66}(1-x)^{0.66}\,, \label{eqrhoP}\\
\phi_{\phi_\|}(x;\zeta_2) &=& 3.14\, x^{0.64}(1-x)^{0.64}\,,\\
\phi_{\rho_\bot}(x;\zeta_2) &=&2.73\, x^{0.49}(1-x)^{0.49}\,, \label{eqrhoT} \\
\phi_{\phi_\bot}(x;\zeta_2) &=& 2.64\, x^{0.48}(1-x)^{0.48}\,.
\end{eqnarray}
\end{subequations}

In order to demonstrate the accuracy of these reconstructions we report that, when working with \mbox{$j_{\rm max}=2$}, then all exponents in Eqs.\,\eqref{phiresults} are unchanged and
\begin{equation}
\begin{array}{ll}
a_{2,\rho_\|} = -0.0017\,, &
a_{2,\phi_\|} = -0.0020\,, \\
a_{2,\rho_\perp} = \phantom{-} 0.0014\,, &
a_{2,\phi_\perp} = \phantom{-} 0.0017\,.
\end{array}
\end{equation}
Plainly, nothing is gained by considering $a_{j\geq 4}$ in Eq.\,\eqref{PDAGalpha}.

The pion PDA depicted in Figs.\,\ref{fig:rda} and \ref{fig:comp} is the RL result from Ref.\,\cite{Chang:2013pq}:
\begin{equation}
\label{resphipi2}
\varphi_\pi^{\rm RL}(x) = 1.74 [x (1-x)]^{\alpha_-^\pi} \, [1 + a_2^\pi C_2^{\alpha_\pi}(2 x - 1)]\,.
\end{equation}
with $\alpha_{\pi} = 0.79$, $a_2^{\pi}=0.0029$, which may be represented as $\varphi_\pi^{\rm RL}(x) \approx 1.71 [x(1-x)]^{0.28}$.

The advantages of the ``Gegenbauer-$\alpha$'' procedure are further highlighted by projecting the $\rho$-meson results above onto the conformal expansion in Eq.\,\eqref{PDAG3on2}.  Using Eq.\,\eqref{projection}, one finds:
\begin{equation}
a^{3/2}_{2\|}=0.092\,,\; a^{3/2}_{4\|}=0.031\,,
\; a^{3/2}_{6\|}=0.015\,,
\end{equation}
with $a^{3/2}_{8\|}/a^{3/2}_{2\|}<10$\%;
and
%
\begin{equation}
a^{3/2}_{2\bot}=0.15\,,\;
a^{3/2}_{4\bot}=0.060\,, \;
a^{3/2}_{6\bot}=0.026\,,
\end{equation}
with $a^{3/2}_{10\bot}$ being the first coefficient that is less-than 10\% of $a^{3/2}_{2\bot}$.
Evidently, the conformal expansions converge slowly.  Moreover, the dot-dashed curve in Fig.\,\ref{fig:rda} displays what is typically obtained if one uses limited information about a PDA, such as just one or two of its low-order moments, to constrain the leading coefficient in Eq.\,\eqref{PDAG3on2}; namely, a ``double-humped'' distribution whose pointwise behaviour presents a misleading picture of the true amplitude.

It is evident in Figs.\,\ref{fig:rda} and \ref{fig:comp} and from Eqs.\,\eqref{phiresults}, \eqref{resphipi2} that the PDAs associated with light-quark meson charge-conjugation eigenstates are concave functions whose widths are ordered as follows
\begin{equation}
\label{sizemeson}
\varphi^{\rm asy} <_{N} \varphi_{\rho_\|} <_{N} \varphi_{\phi_\|} <_{N} \varphi_{\rho_\perp} <_{N} \varphi_{\phi_\perp} <_{N} \varphi_{\pi}\,,
\end{equation}
where ``$<_{N}$'' means ``narrower than''.  This result confirms the pattern anticipated in Ref.\,\cite{Segovia:2013eca} following a consideration of meson electric and magnetic charge radii.  It is also notable because, thus far, numerical simulations of lattice-regularised QCD are unable to distinguish between the PDAs associated with $\pi$-, $\rho$- and $\phi$-mesons; i.e., lattice-QCD produces moments of the different distributions that are equal within errors \cite{Braun:2007zr,Arthur:2010xf}.

Equation~\eqref{sizemeson} predicts an ordering of valence-quark light-front spatial-extent within mesons; viz., this extent is smallest within the pion and increases through the $\perp$-polarisation to the $\|$-polarisation.
We choose to quantify this via $r_{\rm LF}$, where
\begin{equation}
r_{\rm LF}^2 = -\left. \varphi^{\prime\prime}(x) \right|_{x=1/2}\,;
\end{equation}
and in this way one finds, relative to $r_{\rm LF}^{\varphi^{\rm asy}}$,
\begin{equation}
\label{radii}
\begin{array}{ccccc}
r_{\rm LF}^{\pi} & r_{\rm LF}^{\phi_\perp} & r_{\rm LF}^{\rho_\perp} & r_{\rm LF}^{\phi_\|} & r_{\rm LF}^{\rho_\|} \\
0.467 \; & 0.636 \; & 0.644 \; & 0.756  \; & 0.771
\end{array}\,.
\end{equation}
%
%
%
Effects induced by SU$(3)$-flavour symmetry breaking are significantly smaller than those connected with altering the light-front polarisation of vector mesons.  These observations are consistent with: the simpler $\pi$-meson Bethe-Salpeter amplitude supporting less repulsion from spin-orbit-like interactions than the more complicated vector meson amplitude; and the violation of SU$(3)$-flavour symmetry being modulated primarily by mass-scales associated with dynamical instead of explicit chiral symmetry breaking ($f_K/f_\pi = 1.2$ cf.\ $\hat m_s/\hat m_{u/d} = 22$).
%

Equation~\eqref{radii} actually signals an important feature of our results for the distribution amplitudes; namely, their dilation with respect to $\varphi^{\rm asy}(x)$: each vector meson PDA, although narrower than the pion's PDA computed in the same (RL) truncation, is significantly broader than the asymptotic distribution.  Within the context of QCD, a useful measure of this dilation is the energy scale to which one must evolve a given PDA in order that $\varphi^{\rm asy}$ may be considered a reliable approximation; and, as remarked above, it has long been known \cite{Georgi:1951sr,Gross:1974cs,Politzer:1974fr} that this can only be the case when it is accurate to write $u_{\rm v}(x) \approx \delta(x)$.  For the pion, this situation is only achieved at energy scales $\zeta > \zeta_{\rm LHC}$, where $\zeta_{\rm LHC}$ characterises the energy available at the LHC \cite{Cloet:2013jya}.  Vector meson parton distribution functions have not been measured, so another quantitative criterion should be employed.  A reasonable choice is to judge that $\varphi^{\rm asy}(x)$ is a good approximation to a given PDA at that scale $\zeta_\approx$ for which $a_2^{3/2}(\zeta_\approx)$, computed from the PDA via Eq.\,\eqref{PDAG3on2}, is negligible, where negligible means 5\% or less.  This requirement corresponds to a concave PDA $\propto [x (1-x)]^\alpha$, with $\alpha \gtrsim 0.8$; and, again, it is not uniformly achieved for the vector mesons described by Eqs.\,\eqref{phiresults} unless $\zeta_\approx \gtrsim \zeta_{\rm LHC}$.  [N.B.\ A distribution characterised by $\alpha = 0.8$ would add an entry with value $0.87$ to Eq.\,\eqref{radii}.]

\begin{table}[t]
\caption{Computed moments of the $\rho$-meson PDAs, Eq.\,\eqref{zetam}, compared with selected results obtained elsewhere, using:
AdS/QCD models for light cone wave functions fitted to HERA data \cite{Forshaw:2010py};
QCD sum rules \cite{Ball:1996tb,Ball:1998sk};
nonlocal condensates \cite{Bakulev:1998pf,Pimikov:2013usa};
light-front quantum mechanics \cite{Choi:2007yu};
and
lattice-QCD \cite{Braun:2007zr,Arthur:2010xf}.
We list values obtained with $\varphi = \varphi^{\rm asy}$, Eq.\,\eqref{phiasy}, and $\varphi =\,$constant because they represent lower and upper bounds, respectively, for concave distributions.
\label{Table:moments}
}
\begin{tabular*}
{\hsize}
{
l@{\extracolsep{0ptplus1fil}}
c|@{\extracolsep{0ptplus1fil}}
l@{\extracolsep{0ptplus1fil}}
l@{\extracolsep{0ptplus1fil}}
l@{\extracolsep{0ptplus1fil}}
l@{\extracolsep{0ptplus1fil}}
l@{\extracolsep{0ptplus1fil}}}
 $\langle (2x-1)^m \rangle$    & & $m=2$ & $4$ & $6$ & $8$ & $10$ \\\hline

herein & $\|$ & 0.23 & 0.11 & 0.066 & 0.045 & 0.033 \\
       & $\perp$ & 0.25 & 0.13 & 0.079 & 0.055 & 0.042 \\\hline
$\varphi = \varphi^{\rm asy}$
        & & 0.20 & 0.086 & 0.048 & 0.030 & 0.021 \\
$\varphi =\,$constant
        & & 0.33 & 0.2 & 0.14 & 0.11 & 0.091\\\hline
\cite{Forshaw:2010py} 
       & $\|$ & 0.23 & 0.11 & 0.062 & 0.041 & 0.029\\
       & $\perp$ & 0.26 & 0.13 & 0.079 & 0.054 & 0.039\\\hline%
\cite{Ball:1998sk} 
        & $\|$ & 0.26 &  &   &   &   \\
        & $\perp$ & 0.27 &  &   &   &   \\\hline
\cite{Bakulev:1998pf} 
        & $\|$ & 0.23(1) & 0.095(5)& 0.051(4) &0.030(2)& 0.020(5)\\
        & $\perp$ & 0.33(1) &  &   &   &   \\\hline
\cite{Pimikov:2013usa} 
        & $\|$ & 0.22(2) & 0.089(9)& 0.048(5) &0.030(3)& 0.022(2)\\
        & $\perp$ & 0.11(1) & 0.022(2) &   &   &   \\\hline
\cite{Choi:2007yu} 
       & $\|$ & 0.20(1) & 0.085(5) & 0.045(5) &  &  \\
       & $\perp$ & 0.21(1) & 0.095(5) & 0.05(1) &  &  \\\hline
\cite{Braun:2007zr,Arthur:2010xf} 
    &  & 0.25(2)(2)  &  &   &   &   \\\hline
\end{tabular*}
\end{table}

\subsection{Comparison with other analyses}
It is useful to contrast our predictions with results obtained using a variety of other methods.  One means by which that may be accomplished is to compare calculated values for the moments
\begin{equation}
\label{zetam}
\langle (2x-1)^m \rangle := \int_0^1 dx \, (2x-1)^m\, \phi(x)\,,
\end{equation}
since all tools can at least compute a few low-order moments.  Table~\ref{Table:moments} serves this purpose.

When contemplating Table~\ref{Table:moments}, it is important to be conscious of the fact that moments of a concave distribution must lie between the bounds set by the moments of $\varphi = \varphi^{\rm asy}$, Eq.\,\eqref{phiasy} -- the narrowest achievable distribution, and $\varphi =\,$constant, the broadest possible distribution.  Plainly, the broader that a given distribution is, the closer its moments will align with those obtained from $\varphi =\,$constant.

Evidently, our RL DSE analysis produces vector meson PDAs that are consistent with contemporary simulations of lattice-regularised QCD but broader than all other results except those determined in Ref.\,\cite{Forshaw:2010py}, which were fitted to HERA data on diffractive $\rho$-meson photoproduction \cite{Chekanov:2007zr,Aaron:2009xp}.  In considering Table~\ref{Table:moments}, it should also be borne in mind that only our study and those using lattice-QCD can unambiguously determine the scale at which the calculation is valid; viz., $\zeta=\zeta_2=2\,$GeV.  The model studies, on the other hand, are thought to be defined at some vaguely determined ``typical hadronic scale'', which cannot realistically be known to within better than a factor of two.

\begin{figure}[t]
\centerline{\includegraphics[width=0.42\textwidth]{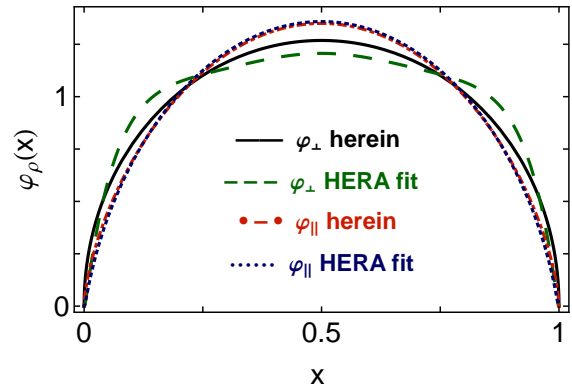}}
\caption{\label{cfForshaw}  Comparison between our predictions for the $\rho$-meson PDAs and those fitted to data on diffractive $\rho$-meson photoproduction in Ref.\,\cite{Forshaw:2010py}.
Curves:
\emph{solid}, $\varphi_\perp(x)$ in Eq.\,\eqref{eqrhoT};
\emph{dashed}, analogue from Ref.\,\cite{Forshaw:2010py};
\emph{dot-dashed}, $\varphi_\|$ in Eq.\,\eqref{eqrhoP};
and \emph{dotted}, analogue from Ref.\,\cite{Forshaw:2010py}.
}
\end{figure}

The agreement between our predictions for the $\rho$-PDA moments and those determined from the data fits in Ref.\,\cite{Forshaw:2010py} is noteworthy.\footnote{Although perhaps a coincidence, it is curious that the DSE prediction for the pion's PDA \cite{Chang:2013pq} is also quantitatively similar to that produced by AdS/QCD models \cite{Brodsky:2006uqa}.}
We therefore illustrate that correspondence further via Fig.\,\ref{cfForshaw}, which displays the complete $x$-dependence of the associated PDAs.  Evidently, there is almost precise pointwise agreement between the respective $\|$-distributions.  Regarding the $\perp$-distributions, the data-fit oscillates mildly around our calculated result.  However, given that adjustments to the fitting form in Ref.\,\cite{Forshaw:2010py} were chosen merely for simplicity, this outcome indicates that the fit may be viewed as approximating our curve: indeed, the lowest six nontrivial moments of both distributions agree within 10\%.  Notably, our prediction yields $m\geq 8$ moments that are larger than those of the data fit; i.e., compared to the fit, $\varphi_{\rho_\perp}(x)$ in Eq.\,\eqref{eqrhoT} actually has more support at the endpoints.  Our predictions for the $\rho$-meson PDAs may therefore be presumed to provide a good, parameter-free description of the HERA data.


\section{Summary and Perspective}
\label{secEpilogue}
We used a rainbow-ladder (RL) truncation of QCD's Dyson-Schwinger equations (DSEs), defined by an interaction that is compatible with modern studies of the gauge sector, in order to compute a parameter-free prediction of $\rho$- and $\phi$-meson valence-quark (twist-two parton) distribution amplitudes  via a light-front projection of their Bethe-Salpeter wave functions.
The PDAs were reconstructed from their moments using a novel technique that provides for rapid convergence.

The PDAs are all broad, concave functions [Eqs.\,\eqref{phiresults}], whose marked dilation with respect to the asymptotic distribution, $\varphi^{\rm asy}(x)=6x(1-x)$, is an expression of dynamical chiral symmetry breaking (DCSB).  Whilst this connection is clearest for the pion, it is also readily apparent for the vector-mesons because: they are constituted from valence-quarks dressed in the same manner as those in the pion; the scales evident in their Bethe-Salpeter amplitudes are fixed by the same dynamics; and vector meson properties, which may reasonably be considered as measuring the scale of DCSB, are large [Table~\ref{tablestatic} and Eqs.\,\eqref{rhoDCSB}].  In addition, there is a clear sense in which the $S$- and $D$-wave components of a light vector-meson's Bethe-Salpeter wave function may be viewed as being of comparable size [Eqs.\,\eqref{ratioN2}, \eqref{SDratio2Real}].

In considering particulars, it is notable that the PDAs can be used to define an ordering of valence-quark light-front spatial-extent within mesons, with this size being smallest within the pion and increasing through the $\perp$-polarisation to the $\|$-polarisation of the vector mesons [Eq.\,\eqref{radii}].  Moreover, effects associated with the breaking of SU$(3)$-flavour symmetry are significantly smaller than those associated with altering the polarisation of vector mesons [Figs.\,\ref{fig:comp}].

Most significantly, perhaps, the pointwise behaviour of our predicted forms for the $\rho$-meson PDAs are in quantitative agreement with the parametrised PDAs fitted recently to data from diffractive vector-meson photoproduction experiments [Fig.\,\ref{cfForshaw}] and can therefore be presumed to provide a good, parameter-free description of that HERA data.

The study described herein may be refined by improving upon the RL truncation; e.g., by using the DCSB-improved gap and Bethe-Salpeter kernels explained in Refs.\,\cite{Chang:2009zb,Chang:2010hb,Chang:2011ei}.  In the pion case, such improvement noticeably softened the distribution \cite{Chang:2013pq}.  We anticipate a qualitatively similar effect on the vector meson PDAs; but it will be weaker because, compared to the pion, vector meson Bethe-Salpeter amplitudes are more complicated and hence more able, even in RL truncation, to express effects such as spin-orbit repulsion, so that their internal structure is less susceptible to modification.  Notwithstanding this, an analysis of the effects of such refinement should be undertaken.

\section*{Acknowledgments}
%
We are grateful for useful comments and observations from
S.\,J.~Brodsky,
I.\,C.~Clo\"et,
R.\,J.~Holt,
J.~Segovia
and
P.\,C.~Tandy.
CDR acknowledges support through an \emph{International Fellow Award} from the Helmholtz Association, and this work was otherwise supported by:
the National Natural Science Foundation of China under Contract Nos.\ 10935001 and 11175004;
the National Key Basic Research Program of China under Contract No.\ G2013CB834400;
the U.\,S.\ Department of Energy, Office of Nuclear Physics, Contract no.~DE-AC02-06CH11357;
and For\-schungs\-zentrum J\"ulich GmbH.

\appendix
\section{Computing renormalisation constants}
\label{Z2ZT}
Owing to the vector and axial-vector Ward identities, one is free to choose the quark wave function renormalisation constant, $Z_2$, as the renormalisation constant for the vector and axial-vector vertices; and that is evident in Eq.\,\eqref{frho}.  This choice guarantees that $f_{\rho,\phi}$ are gauge- and Poincar\'e-invariant, and also independent of the renormalisation point and the regularisation mass-scale.

On the other hand, whilst the values of the tensor couplings $f^T_{\rho,\phi}$ are gauge- and Poincar\'e-invariant, they depend on the renormalisation scale.  In this case, no Ward identity can be used to fully determine $Z_T$ in Eq.\,\eqref{frhoT}.  Instead, it can be computed by solving the inhomogeneous Bethe-Salpeter equation for the dressed tensor vertex,
\begin{equation}
\Gamma_{\mu\nu}(k;P;\zeta)=S_1(k;P;\zeta)\sigma_{\mu\nu}+\ldots\,,
\end{equation}
at zero total momentum, $P=0$.  Then $Z_T(\zeta,\Lambda)$ is the factor required as a multiplier for the Bethe-Salpeter equation inhomogeneity, $\sigma_{\mu\nu}$, in order to achieve $S_1(k^2=\zeta^2;P=0;\zeta) = 1$.

At one-loop order in QCD \cite{Barone:1997fh}:
\begin{equation}
\label{GammaTzeta}
\Gamma_{\mu\nu}(k;P;\zeta) \stackrel{\zeta^2 \gg \Lambda_{\rm QCD}^2}{=} \left[\frac{\alpha_S(\zeta_0^2)}{\alpha_S(\zeta^2)}\right]^{\eta_T}
\Gamma_{\mu\nu}(k;P;\zeta_0)\,,
\end{equation}
where $\eta_T=(-1/3) \gamma_m$.  The pointwise behaviour of $\Gamma_{\mu\nu}(k;P=0;\zeta)$ is illustrated in Ref.\,\cite{Yamanaka:2013zoa}.

Equation~\eqref{GammaTzeta} entails
\begin{equation}
f^T(\zeta) \stackrel{\zeta^2 \gg \Lambda_{\rm QCD}^2}{=} \left[\frac{\alpha_S(\zeta_0^2)}{\alpha_S(\zeta^2)}\right]^{\gamma_T} f^T(\zeta_0)\,,
\end{equation}
and hence that $f^T(\zeta)$ increases as $\zeta$ decreases.

It is worth remarking that, given their ``$\gamma \times \gamma$'' Dirac matrix structure, Eq.\,\eqref{eq:bsa}, one could have anticipated that those terms in the vector meson Bethe-Salpeter amplitude associated with $\{\tau_\nu^j | j=6,7,8 \}$ would play an important role in computing an accurate value for $f^T$ in Eq.\,\eqref{frhoT} and ensuring that it evolves correctly under a change in renormalisation scale.

It is notable, too, that in a quantum mechanical system, a tensor amplitude produced by two fermions must necessarily possess at least one unit of angular momentum, $L$, in the state's rest-frame.  If the two-fermion system possesses $J^{PC}=1^{--}$, then the tensor amplitude will possess $L=0$ and $L=2$ correlations because $L=1$ is forbidden by parity conservation.

\section{Gap and Bethe-Salpeter equations}
\label{sec:kernels}
The gap equation in QCD is
\begin{subequations}
\label{gendseN}
\begin{eqnarray}
S_f(p)^{-1} &=& Z_2 \,(i\gamma\cdot p + m_f^{\rm bm}) + \Sigma_f(p)\,,\\
\nonumber \Sigma_f(p) &=& Z_1 \int^\Lambda_{dq}\!\! g^2 D_{\mu\nu}(p-q)\frac{\lambda^a}{2}\gamma_\mu \\
&&\rule{2em}{0ex}\times S_f(q) \frac{\lambda^a}{2}\Gamma^f_\nu(q,p) ,
\end{eqnarray}
\end{subequations}
where: $f$ is a quark flavour label, $D_{\mu\nu}$ is the gluon propagator; $\Gamma^f_\nu$, the quark-gluon vertex; $\int^\Lambda_{dq}$, a symbol that represents a Poincar\'e invariant regularisation of the four-dimensional Euclidean integral, with $\Lambda$ the regularisation mass-scale (a Pauli-Villars-like scheme is usually adequate, see, e.g., Refs.\,\protect\cite{Holl:2005vu,Chang:2008ec}); $m_f^{\rm bm}(\Lambda)$, the current-quark bare mass; and $Z_{1,2}(\zeta^2,\Lambda^2)$, respectively, the vertex and quark wave-function renormalisation constants, with $\zeta$ the renormalisation point.  Regarding renormalisation, we follow precisely the procedures of Refs.\,\cite{Maris:1997tm,Maris:1999nt} and work with $\zeta=2\,$GeV, which is a scale typical of contemporary numerical simulations of lattice-regularised QCD.  The solution of Eq.\,\eqref{gendseN} has the form
\begin{equation}
\label{SpDressed}
S_f(p) = -i \gamma\cdot p \, \sigma_V^f(p^2) + \sigma_S^f(p^2)\,.
\end{equation}

In rainbow-ladder truncation the model input is expressed in a statement about the nature of the gap equation's kernel at infrared momenta, since the behaviour at momenta $k^2\gtrsim 2\,$GeV$^2$ is fixed by perturbation theory and the renormalisation group \cite{Jain:1993qh,Maris:1997tm}.  In Eq.\,\eqref{gendseN}, this amounts to writing ($k=p-q$)
\begin{eqnarray}
\nonumber
\lefteqn{
Z_1 g^2 D_{\mu\nu}(k) \Gamma_\nu(q,p) = k^2 {\cal G}(k^2)
D^{\rm free}_{\mu\nu}(k) \gamma_\nu Z_2^2}\\
&=&   \left[ k^2 {\cal G}_{\rm IR}(k^2) + 4\pi \tilde\alpha_{\rm pQCD}(k^2) \right]
D^{\rm free}_{\mu\nu}(k) \gamma_\nu\, Z_2^2 ,
\label{rainbowdse}
\end{eqnarray}
wherein $D^{\rm free}_{\mu\nu}(k)$ is the Landau-gauge free-gauge-boson propagator;
$\tilde\alpha_{\rm pQCD}(k^2)$ is a bounded, monotonically-decreasing regular continuation of the perturbative-QCD running coupling to all values of spacelike-$k^2$; and ${\cal G}_{\rm IR}(k^2)$ is an \emph{Ansatz} for the interaction at infrared momenta: ${\cal G}_{\rm IR}(k^2)\ll \tilde\alpha_{\rm pQCD}(k^2)$ $\forall k^2\gtrsim 2\,$GeV$^2$.  The form of ${\cal G}_{\rm IR}(k^2)$ determines whether confinement and/or DCSB are realised in solutions of the gap equation.  

\begin{table}[tb]
\caption{Parameters associated with Eq.\,\eqref{Spfit}, which express our interpolation of the $u/d$- and $s$-quark propagators.  (Dimensioned quantities in GeV.)
\label{Table:Sparameters}
}
\begin{tabular*}
{\hsize}
{
c|@{\extracolsep{0ptplus1fil}}
c@{\extracolsep{0ptplus1fil}}
c@{\extracolsep{0ptplus1fil}}
c@{\extracolsep{0ptplus1fil}}
c@{\extracolsep{0ptplus1fil}}
c@{\extracolsep{0ptplus1fil}}}
      & $z_1$ & $m_1$  & $z_s$ & $m_2$ \\\hline
$u/d$ & $(0.40,0.015)$ & $(0.56,0.20)$ & $(0.18,0)$ & $(-1.30,-0.60)$ \\
  $s$ & $(0.34,0.2)$ & $(0.80,0.20)$ &  &  \\\hline
\end{tabular*}
\end{table}

The interaction in Ref.\,\cite{Qin:2011dd} is
\begin{equation}
\label{CalGQC}
{\cal G}(s) = \frac{8 \pi^2}{\omega^4} D \, {\rm e}^{-s/\omega^2}
+ \frac{8 \pi^2 \gamma_m\, {\cal F}(s)}{\ln [ \tau + (1+s/\Lambda_{\rm QCD}^2)^2]} ,
\end{equation}
where: $\gamma_m = 12/(33-2 N_f)$, $N_f=4$, $\Lambda_{\rm QCD}=0.234\,$GeV; $\tau={\rm e}^2-1$; and ${\cal F}(s) = \{1 - \exp(-s/[4 m_t^2])\}/s$, $m_t=0.5\,$GeV.
With $D\omega = \,$constant, light-quark observables are independent of the value of $\omega \in [0.4,0.6]\,$GeV.  We use $\omega =0.5\,$GeV and $D\omega = (0.87\,{\rm GeV})^3$, with which a renormalisation point invariant current-quark mass $\hat m_{u/d}=6.4\,$MeV produces $m_\pi=0.14\,$GeV and $f_\pi=0.092\,$GeV.

\begin{table}[tb]
\centering
\caption{Fit parameters for the vector meson Bethe-Salpeter amplitudes. \label{tab:prmvector}}
\begin{tabular*}
{\hsize}
{
c|@{\extracolsep{0ptplus1fil}}
r@{\extracolsep{0ptplus1fil}}
r@{\extracolsep{0ptplus1fil}}
r@{\extracolsep{0ptplus1fil}}
c@{\extracolsep{0ptplus1fil}}
c@{\extracolsep{0ptplus1fil}}
l@{\extracolsep{0ptplus1fil}}
l@{\extracolsep{0ptplus1fil}}}
\hline
$\rho$ & $c^{\rm i}$ & $c^{u}$ & $\phantom{-}\nu^{\rm i}$ & $\nu^{\rm u}$ & $a$\phantom{00} & $\Lambda^{\rm i}$ & $\Lambda^{\rm u}$ \\
\hline
${\cal F}_1$&1.38&0.013&-0.87&1&2.5&1.3&0.6\\
${\cal F}_2$&0.14&0.004&-0.83&1&$6/[\Lambda_{{\cal F}_2}^{\rm i}]^2$&1.1&0.65\\
${\cal F}_3$&-1.50&-0.013&-0.80&1&$6.5/[\Lambda_{{\cal F}_3}^{\rm i}]^4$&1.15&0.77\\
${\cal F}_4$&-0.61&-0.004&0.15&1&$2.7/[\Lambda_{{\cal F}_4}^{\rm i}]^2$&1.05&0.62\\
${\cal F}_5$&-1.73&-0.003&0.50&1&$2.3/[\Lambda_{{\cal F}_5}^{\rm i}]$&1.15&0.57\\
${\cal F}_6$&0.60&0.020&-0.85&1&$2.3/[\Lambda_{{\cal F}_6}^{\rm i}]^3$&1.3&0.5\\
${\cal F}_7$&0.10&-0.004&-0.83&1&$5.8/[\Lambda_{{\cal F}_7}^{\rm i}]$&1.2&0.45\\
${\cal F}_8$&0.20&0.013&-0.87&1&$5.5/[\Lambda_{{\cal F}_8}^{\rm i}]$&1.4&0.57\\
\hline
\end{tabular*}

\bigskip

\begin{tabular*}
{\hsize}
{
c|@{\extracolsep{0ptplus1fil}}
r@{\extracolsep{0ptplus1fil}}
r@{\extracolsep{0ptplus1fil}}
r@{\extracolsep{0ptplus1fil}}
c@{\extracolsep{0ptplus1fil}}
c@{\extracolsep{0ptplus1fil}}
l@{\extracolsep{0ptplus1fil}}
l@{\extracolsep{0ptplus1fil}}}
\hline
$\phi$ & $c^{\rm i}$ & $c^{u}$ & $\phantom{-}\nu^{\rm i}$ & $\nu^{\rm u}$ & $a$\phantom{00} & $\Lambda^{\rm i}$ & $\Lambda^{\rm u}$ \\
\hline
${\cal F}_1$&1.08&0.009&-0.79&1&2.4&1.5&0.9\\
${\cal F}_2$&0.14&0.001&-0.85&1&$6.5/[\Lambda_{{\cal F}_2}^{\rm i}]^2$&1.05&0.94\\
${\cal F}_3$&-1.50&-0.007&-0.82&1&$6.8/[\Lambda_{{\cal F}_3}^{\rm i}]^4$&1.1&0.95\\
${\cal F}_4$&-0.81&-0.003&0.20&1&$2.75/[\Lambda_{{\cal F}_4}^{\rm i}]^2$&1.35&0.9\\
${\cal F}_5$&-1.63&-0.008&0.40&1&$2.1/[\Lambda_{{\cal F}_5}^{\rm i}]$&1.45&0.89\\
${\cal F}_6$&0.10&-0.050&-0.78&1&$2.5/[\Lambda_{{\cal F}_6}^{\rm i}]^3$&1.4&0.3\\
${\cal F}_7$&0.08&-0.006&-0.74&1&$5.7/[\Lambda_{{\cal F}_7}^{\rm i}]$&1.2&0.4\\
${\cal F}_8$&0.12&0.002&-0.82&1&$5.3/[\Lambda_{{\cal F}_8}^{\rm i}]$&1.4&0.37\\

\hline
\end{tabular*}
\end{table}

In rainbow-ladder truncation, in the isospin symmetric limit, the vector meson Bethe-Salpeter equation is
\begin{equation}
\Gamma_{\nu}(k;P) =
 - \frac{4}{3}Z_2^2\int_{dq}^\Lambda {\cal G}(k-q)\, D^{\rm free}_{\alpha\beta}(k-q)\, \gamma_\alpha  \chi_{\nu}(q;P) \gamma_\beta\,,
\end{equation}
where $\chi_{\nu}(q;P)$ is the Bethe-Salpeter wave function defined in Eq.\,\eqref{BSwavefunction}.  We use the following complete, orthogonal tensor basis \cite{Maris:1999nt}:
{\allowdisplaybreaks
\begin{subequations}
\label{eq:bsa}
\begin{eqnarray}
\tau_\nu^1(q,P)&=&i\gamma^T_\nu\,, \\
\tau_\nu^2(q,P)&=&i[3q^T_\nu(\gamma\cdot q^T)-\gamma^T_\nu(q^T)^2]\,, \\
\tau_\nu^3(q,P)&=&i(P\cdot q)q^T_\nu \gamma\cdot P \,,  \\
\tau_\nu^4(q,P)&=&i[\gamma^T_\nu \gamma\cdot P(\gamma\cdot q^T)+q^T_\nu \gamma\cdot P]\,, \\
\tau_\nu^5(q,P)&=&q^T_\nu\,, \\
\tau_\nu^6(q,P)&=&(q\cdot P)[\gamma^T_\nu(\gamma\cdot q^T)-(\gamma\cdot q^T)\gamma^T_\nu]\,, \\
%
\tau_\nu^7(q,P)&=& \gamma^T_\nu \gamma\cdot P - \gamma\cdot P \gamma^T_\nu
- 2 \tau_\nu^8(q,P)\,,\\
%
\tau_\nu^8(q,P)&=&\hat q^T_\nu(\gamma\cdot \hat q^T) \gamma\cdot P\,,
\end{eqnarray}
\end{subequations}}
\hspace*{-0.5\parindent}where $P\cdot a^T =0$ for any four-vector $a_\mu$ and $\hat q^T \cdot\hat q^T=1$.

N.B.\, The factors of $Z_2^2$ here and in Eq.\,\eqref{rainbowdse} ensure multiplicative renormalisability \cite{Bloch:2002eq}.

\section{Interpolations of propagators and Bethe-Salpeter amplitudes}
\label{NakanishiAppendix}
Here we describe the interpolations used in our evaluation of the moments in Eq.\,\eqref{momentsE}.  The dressed-quark propagators are represented as \cite{Bhagwat:2002tx}
\begin{equation}
S(p) = \sum_{j=1}^{j_m}\bigg[ \frac{z_j}{i \gamma\cdot p + m_j}+\frac{z_j^\ast}{i \gamma \cdot p + m_j^\ast}\bigg], \label{Spfit}
\end{equation}
with $\Im m_j \neq 0$ $\forall j$, so that $\sigma_{V,S}$ are meromorphic functions with no poles on the real $p^2$-axis, a feature consistent with confinement \cite{Roberts:2007ji,Chang:2011vu,Bashir:2012fs,Cloet:2013jya}.  We find that $j_m=2$ is adequate for $u/d$-quarks and that $j_m=1$ is satisfactory for the $s$-quark.  The parameter values are given in Table~\ref{Table:Sparameters}.

We retain all eight terms in the vector meson Bethe-Salpeter amplitude, Eq.\,\eqref{rhoBSA} and, with relative momentum defined by $\eta=1/2$, we fit the associated scalar functions via
{\allowdisplaybreaks
\begin{subequations}
\begin{eqnarray}
{\cal F}_j(q;P) &=& {\cal F}_j^{\rm i}(q;P) + {\cal F}_j^{\rm u}(q;P) \,, \; \\
\nonumber {\cal F}_j^{\rm i}(q;P) & = &
c_{{\cal F}_j}^{\rm i}
\int_{-1}^1 \! dz \, \rho_{\nu^{\rm i}_{{\cal F}_j}}(z) \bigg[
a_{\cal F} \hat\Delta_{\Lambda^{\rm i}_{{\cal F}_j}}^4(q_z^2) \\
&& \rule{7em}{0ex}
+ a^-_{\cal F} \hat\Delta_{\Lambda^{\rm i}_{{\cal F}_j}}^5(q_z^2)
\bigg], \label{Fifit}\\
{\cal F}_k^{\rm u}(q;P) & = & c_{{\cal F}_k}^{\rm u} \int_{-1}^1 \! dz \, \rho_{\nu^{\rm u}_{{\cal F}_k}}(z)
\frac{1}{[\Lambda_{{\cal F}_k}^{\rm i} ]^{{\mathpzc n}_k}}
\hat\Delta^{\mathpzc{p}_k}_{\Lambda^{\rm u}_{{\cal F}_k}}(q_{z_k}^2)\,, \quad\quad
\end{eqnarray}
\end{subequations}}
\hspace*{-0.5em}with:
${\mathpzc n}_1 = 0$,
${\mathpzc n}_{2,4} = 2$,
${\mathpzc n}_3 = 4$,
${\mathpzc n}_{5,7,8} = 1$,
${\mathpzc n}_{6} = 3$;
${\mathpzc p}_{1,7,8}=1$,
${\mathpzc p}_{2,\ldots,6}=2$;
and, for $k=1,\ldots,8$,
\begin{equation}
a^-_{{\cal F}_k} = 1/[\Lambda_{{\cal F}_k}^{\rm i} ]^{{\mathpzc n}_k} - a_{{\cal F}_k},
\end{equation}
$q_{z_k}^2=q^2 + z q\cdot P + (q\cdot p)^2/(\Lambda^{\rm u}_{{\cal F}_k})^2$.  The parameters obtained for the $\rho$ and $\phi$ mesons are listed in Tables~\ref{tab:prmvector}.

\end{document}